\documentclass[preprint2]{aastex}

\slugcomment{Accepted to ApJ}

\shorttitle{NGC 3766}
\shortauthors{McSwain et al.}

\begin{document}

\title{The B and Be Star Population of NGC 3766}

\author{M.\ Virginia McSwain\altaffilmark{1,2,3}}
\affil{Department of Astronomy, Yale University, P.O.\ Box 208101, New 
Haven, CT 06520-8101; mcswain@lehigh.edu}

\author{Wenjin Huang}
\affil{Department of Astronomy, California Institute of Technology, 
MC 105-24, Pasadena, CA 91125; wenjin@astro.caltech.edu}

\author{Douglas R.\ Gies, Erika D.\ Grundstrom\altaffilmark{4}}
\affil{Center for High Angular Resolution Astronomy, Department of Physics and Astronomy, Georgia State University, P.O.\ Box 4106, Atlanta, GA 30302-4106; gies@chara.gsu.edu, erika@chara.gsu.edu}

\author{Richard H.\ D.\ Townsend}
\affil{Bartol Research Institute, University of Delaware, Newark, DE 19716; rhdt@bartol.udel.edu}

\altaffiltext{1}{Visiting Astronomer, Cerro Tololo Inter-American Observatory.
CTIO is operated by AURA, Inc.\ under contract to the National Science
Foundation.}
\altaffiltext{2}{NSF Astronomy and Astrophysics Postdoctoral Fellow}
\altaffiltext{3}{Current address: Department of Physics, Lehigh University, 16 Memorial Drive East, Bethlehem, PA 18015}
\altaffiltext{4}{Current address: Physics and Astronomy Department, Vanderbilt University, 1807 Station B, Nashville, TN 37235}

\begin{abstract}

We present multiple epochs of H$\alpha$ spectroscopy for 47 members of the open cluster NGC 3766 to investigate the long term variability of its Be stars.  Sixteen of the stars in this sample are Be stars, including one new discovery.  Of these, we observe an unprecedented 11 Be stars that undergo disk appearances and/or near disappearances in our H$\alpha$ spectra, making this the most variable population of Be stars known to date.  NGC 3766 is therefore an excellent location to study the formation mechanism of Be star disks.  From blue optical spectra of 38 cluster members and existing Str\"omgren photometry of the cluster, we also measure rotational velocities, effective temperatures, and polar surface gravities to investigate the physical and evolutionary factors that may contribute to the Be phenomenon.  Our analysis also provides improvements to the reddening and distance of NGC 3766, and we find $E(B-V) = 0.22 \pm 0.03$ and $(V-M_{\rm V})_0 = 11.6 \pm 0.2$, respectively.  The Be stars are not associated with a particular stage of main-sequence evolution, but they are a population of rapidly rotating stars with a velocity distribution generally consistent with rotation at $70-80$\% of the critical velocity, although systematic effects probably underestimate the true rotational velocities so that the rotation is much closer to critical.  Our measurements of the changing disk sizes are consistent with the idea that transitory, nonradial pulsations contribute to the formation of these highly variable disks.  

\end{abstract}


\section{Introduction}

\setcounter{footnote}{3}
NGC 3766 is a rich, young open cluster in the Carina spiral arm that is
well known for its high content of Be stars \citep{slettebak1985}, and
many previous studies of this cluster have focused on the
characteristics of these stars to identify their evolutionary status. 
The cluster has been the target of numerous photometric studies
\citep{ahmed1962, yilmaz1976, shobbrook1985, shobbrook1987,
moitinho1997, piatti1998, tadross2001, mcswain2005b}.  But despite these
intensive investigations, the cluster's age and distance remain somewhat
uncertain; measurements of its age range from 14.5 to 25 Myr
(WEBDA\footnote{The WEBDA database is maintained by E.\ Paunzen and is
available online at http://www.univie.ac.at/webda/navigation.html.};
\citealt{lynga1987, moitinho1997, tadross2001}), and its distance is
between 1.5 and 2.2 kpc.  The reddening $E$($B$-$V$) is between 0.16 and
0.22 (see the discussion of \citealt{moitinho1997}). 

Spectroscopic investigations of NGC 3766 have targeted a limited sample
of cluster members, focusing primarily on the Be star and supergiant
populations (\citealt{harris1976}; \citealt{mermilliod1982} and
references therein; \citealt{slettebak1985, levesque2005}).  Even the
eclipsing double-lined spectroscopic binary BF Centauri (= HD 100915), a
member of NGC 3766, has been largely neglected by modern spectroscopic
observations (\citealt{clausen2007} and references therein).  For most
cluster members, no detailed information about their physical
characteristics such as temperature, gravity, rotation, and metallicity
are known. 

In this work, we present red and blue optical spectra for both normal
B-type and Be stars in the cluster.  Like many prior studies of NGC
3766, our primary goal is to investigate the Be star population; but
unlike other works, we achieve a more complete understanding of this
subset of B stars by comparing these emission-line objects to their
non-emission counterparts.  Therefore we present measurements of the
effective temperature, $T_{\rm eff}$, surface gravity, $\log g$, and in
most cases the projected rotational velocity, $V \sin i$, for 26 normal
B stars and 16 Be stars in NGC 3766.  We use these results to improve
the known reddening and distance to the cluster.  From multiple epochs
of H$\alpha$ spectroscopy, we also investigate the variability of the
circumstellar disks and estimate the disk mass loss/gain rates for 11 Be
stars.  Finally, we use the observed disk masses and angular momenta to
show that nonradial pulsations are a possible origin for the disks, and
they probably fill during short-lived bursts of mass flow from the
stellar surface.

%%%%%%%%%%%%%%%%%%%%%%%%%%%%%%%%%%%%%%%%%%%%%

\section{Observations}

We obtained spectra of NGC 3766 during multiple observing runs in 2003
March, 2005 February, 2006 May, and 2007 January$-$July using the CTIO
Blanco 4-m telescope with the Hydra multi-fiber spectrograph and the
CTIO 1.5-m telescope with the Cassegrain spectrograph, operated by the
SMARTS Consortium.  The details of all runs are summarized in Table
\ref{spectroscopy}.  Most of the runs targeted the H$\alpha$ emission
line profile with low spectral resolution to characterize the Be stars'
emission; however, during one run we observed the blue optical region
with higher resolving power to observe numerous other H Balmer and
\ion{He}{1} line profiles and measure the physical parameters of the
cluster members.

We selected the targets for each run by giving highest priority to the
known Be stars in this cluster (save WEBDA No.\ 232, which was saturated
in our photometric study).  We then selected other B-type stars in the
cluster by ranking them according to their $y$--H$\alpha$ color to
preferentially select any weak emission stars that were not detected in
our photometry \citep{mcswain2005a, mcswain2005b}.  All observations
were performed by M.\ V.\ McSwain except the 2007 CTIO 1.5-m runs, which
were taken in service mode by a SMARTS observer.  For the Hydra
observations, we generally began by taking short exposures and then
parking the fibers used for the brightest stars to avoid saturation in
the longer exposures.  Not all of the known Be stars could be observed
in one fiber configuration, so we took three to four exposures each of
two configurations to observe all of the targets.  Therefore up to eight
exposures of each star were obtained with the Hydra runs.  We also
observed a HeNeAr comparison lamp source just before and after the set
of cluster observations for wavelength calibrations.  For the CTIO 1.5-m
observations, we alternated each stellar observation with a Ne
comparison lamp spectrum. 

The CTIO 1.5-m spectra were reduced and rectified to a unit continuum
using standard routines for slit spectra in IRAF.  All of the Hydra
spectra were zero corrected using standard routines in IRAF, and they
were flat fielded, wavelength calibrated, and sky subtracted in IRAF
using the \textit{dohydra} routine.  In comparing the slit spectra and
fiber spectra for many of the same objects in our data set, we find no
evidence of systematic differences in the background subtraction due to
cluster nebulosity. For each Hydra spectral configuration, we
transformed the observations to a common heliocentric wavelength grid
and co-added them to achieve good S/N for each star.

\begin{figure}
\includegraphics[angle=0,scale=.7]{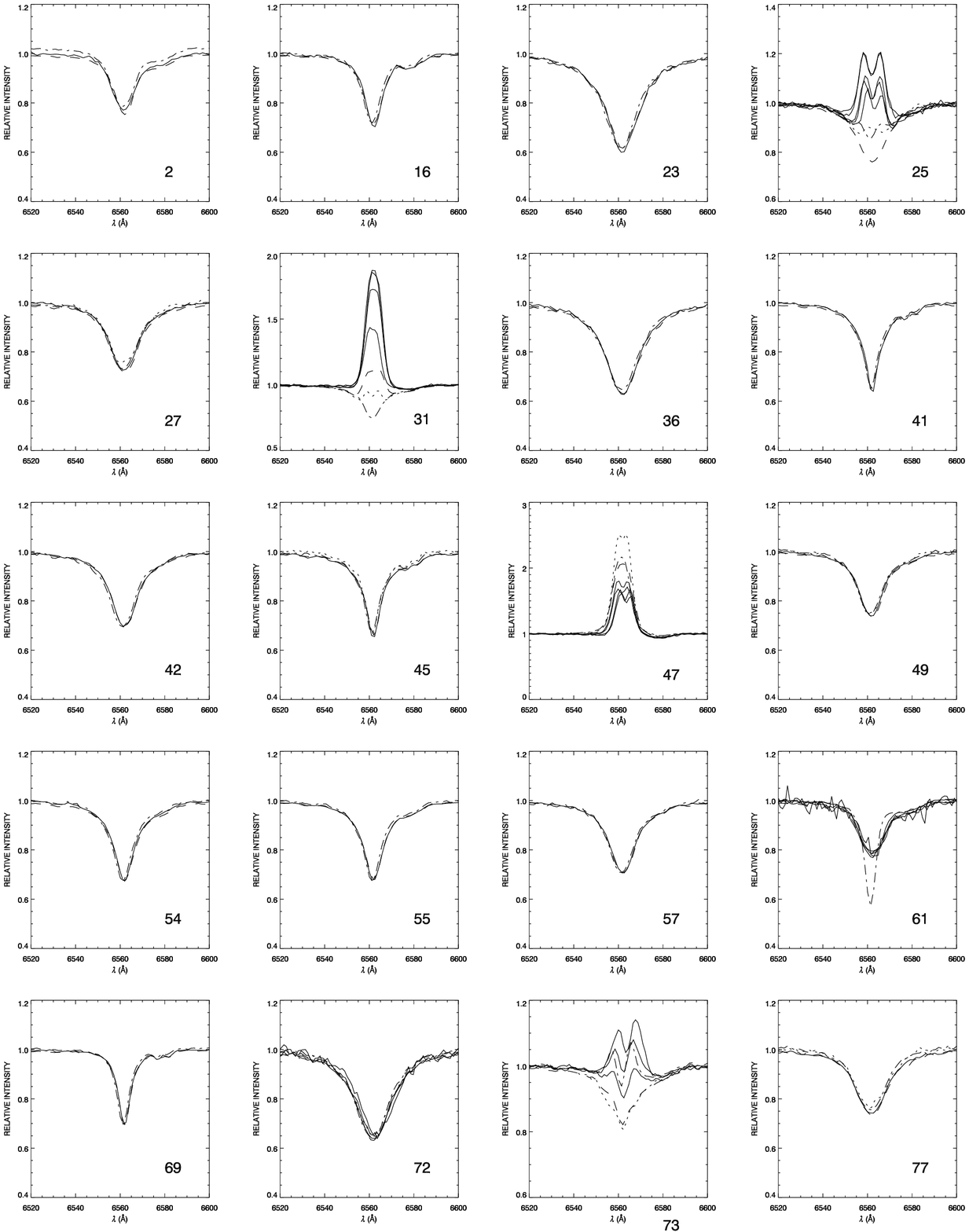}
\caption{H$\alpha$ profiles, labeled by MG ID number.  Spectra from 2003, where available, are shown with \textit{dotted lines}, spectra from 2005 are shown with \textit{dot-dashed lines}, spectra from 2006 are shown with \textit{dashed lines}, and spectra from 2007 are shown with \textit{solid lines}.
\label{specvar1}
} 
\end{figure}

\begin{figure}
\includegraphics[angle=0,scale=.7]{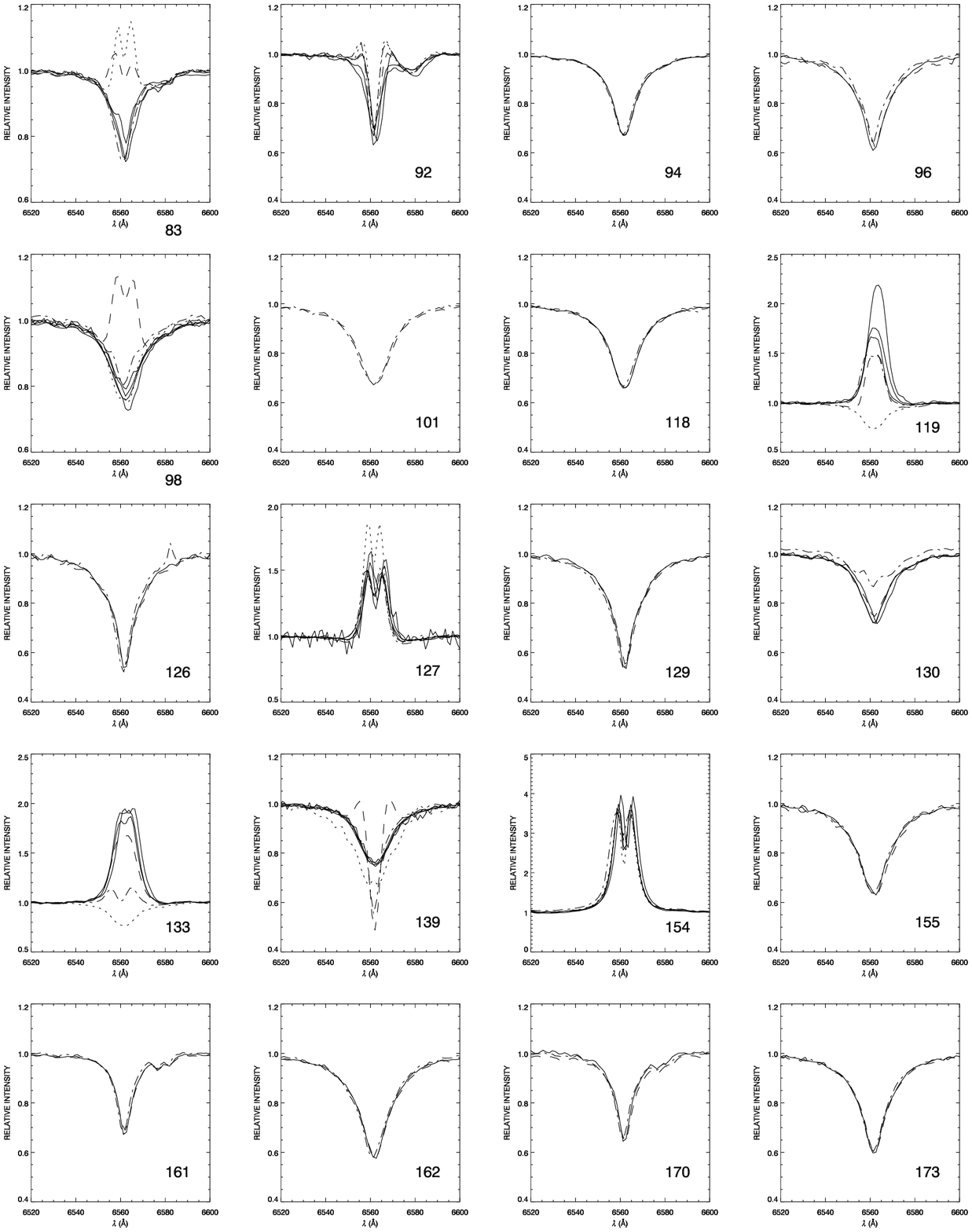}
\caption{H$\alpha$ profiles in the same format as Fig.\ \ref{specvar1}. 
\label{specvar2}
}
\end{figure}

\begin{figure}
\includegraphics[angle=0,scale=.7]{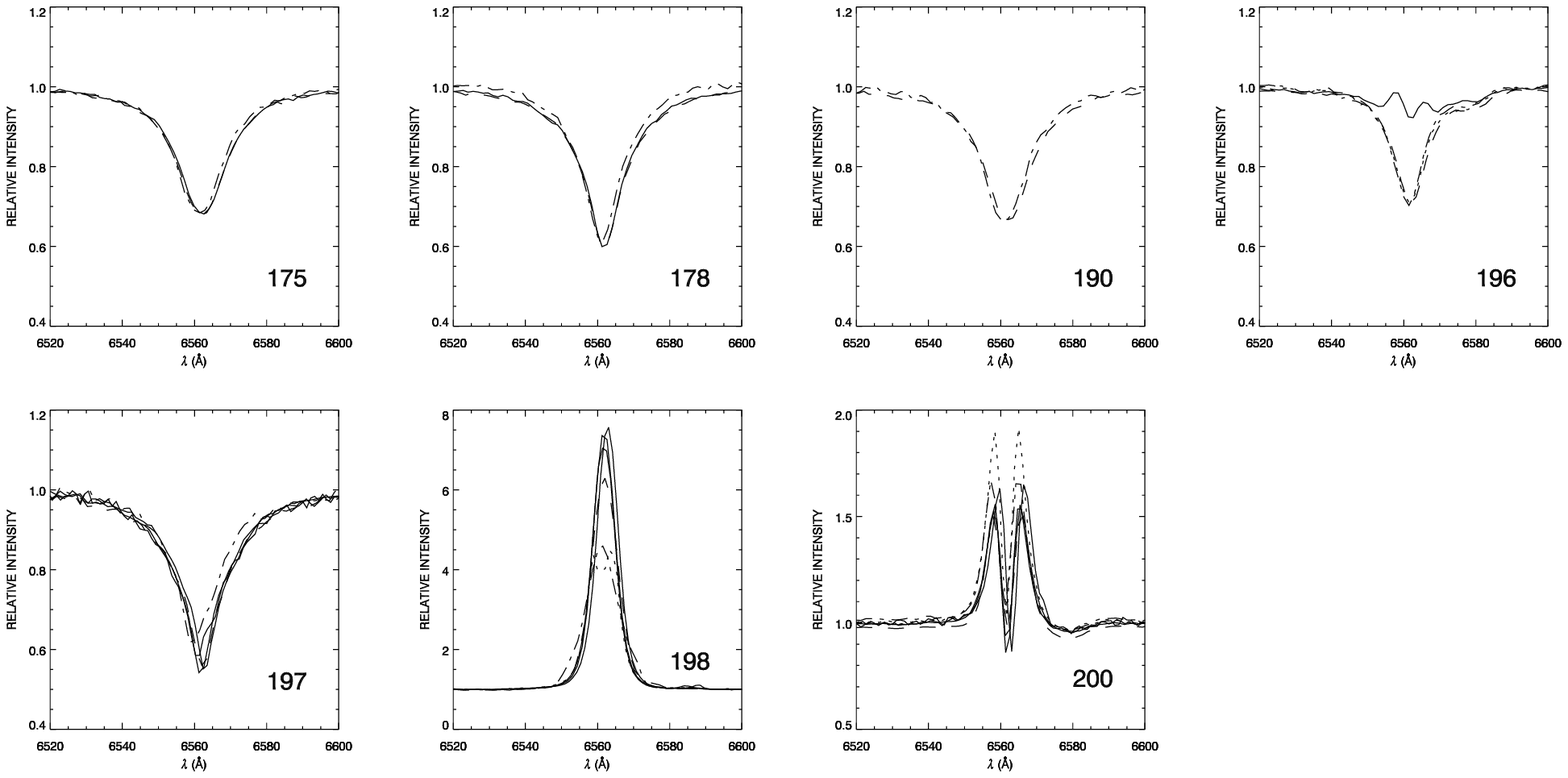}
\caption{H$\alpha$ profiles in the same format as Fig.\ \ref{specvar1}.
\label{specvar3}
}
\end{figure}

The complete sample of stars presented in this work is listed in Table
\ref{eqwidth}.  Column 1 gives each star's identification number based
on the assigned number in \citet{mcswain2005b}; the corresponding WEBDA
numbers are given in column 2, where available.  We obtained H$\alpha$
spectra for each of these stars during at least three epochs in most
cases, and these are shown in Figures \ref{specvar1}--\ref{specvar3}.

\placetable{eqwidth}
\placefigure{specvar1}
\placefigure{specvar2}
\placefigure{specvar3}

%%%%%%%%%%%%%%%%%%%%%%%%%%%%%%%%%%%%%%%%%%

\section{Physical Parameters from Spectral Models}

\subsection{$V \sin i$ Measurements}

We began our investigation of each star's physical parameters by
generating a grid of synthetic, plane-parallel, local thermodynamic
equilibrium (LTE) atmospheric models using the Kurucz ATLAS9 code
\citep{kurucz1994}.  We adopted solar abundances and a microturbulent
velocity of 2 km~s$^{-1}$ for these stars, which corresponds to the mean
microturbulence observed among late-type, main-sequence (MS) B stars
\citep*{lyubimkov2004}.  Each atmospheric model was then used to
calculate a grid of model spectra using SYNSPEC \citep{lanz2003}.

For the 38 stars with available blue spectra, we made a preliminary
estimate of their effective temperature and gravity, $T_{\rm eff}$ and
$\log g$ respectively, by comparing the observed H$\gamma$, H$\delta$,
\ion{He}{1} $\lambda4143$, and \ion{He}{1} $\lambda4471$+\ion{Mg}{2}
$\lambda4481$ line profiles to our grid of Kurucz spectral models.  To
measure $V \sin i$, we compared the observed \ion{He}{1} line profiles
to the model profiles convolved with a limb-darkened, rotational
broadening function and a Gaussian instrumental broadening function.  We
determined the best fit over a grid of values, spaced 2~km~s$^{-1}$
apart, minimizing the mean square of the deviations, rms$^2$.  The
formal error, $\Delta V \sin i$, is the offset from the best-fit value
that increases the rms$^2$ by $2.7 \, \rm rms^2$/$N$, where $N$ is the
number of wavelength points within the fit region.  Our measured $V \sin
i$ and $\Delta V \sin i$ are listed in columns 3--4 of Table
\ref{params}.

Even the \ion{He}{1} lines may contain some weak emission in Be stars,
partially filling and narrowing their line profiles.  Furthermore, a
number of the Be stars show evidence of narrow ``shell'' line components
(formed in the outer disk), and the presence of a shell component may
make the profile appear too narrow in some cases.  Therefore we consider
our $V \sin i$ measurements for the Be stars to be lower limits. 
However, we note that the \ion{He}{1} lines do not exhibit obvious signs
of emission among most of the Be stars, and these lines are much less
susceptible to emission than the H$\gamma$ or H$\delta$ lines.  For the
case of No.\ 154, a shell star with strong emission and contamination
present in the \ion {He}{1} lines, we used the \ion{Mg}{2} $\lambda4481$
line to measure $V \sin i$.

There are few previous measurements of $V \sin i$ for members of NGC
3766 in the literature, but we found that 10 stars in our sample were
also measured by \citet{slettebak1985}.  Our $V \sin i$ measurements
generally agree well, with the exception of No.\ 154.  Slettebak found
$V \sin i = 220$~km~s$^{-1}$ for that star, nearly double our measured
value.  We emphasize that the exceptionally strong He and metal lines of
this shell star amplify the difficulty of measuring its $V \sin i$.

\subsection{$T_{\rm eff}$ and $\log g$ Measurements of B stars}

For the B stars with $T_{\rm eff} < 15,000$ K, we used the ``virtual
star'' method of \citet{huang2006b} to improve our $T_{\rm eff}$ and
$\log g$ measurements.  (Their virtual star is a spherically symmetric
star with constant $T_{\rm eff}$ and $\log g$ across the stellar
surface.)  They generated detailed H$\gamma$ line profiles using
line-blanketed, LTE Kurucz ATLAS9 and SYNSPEC codes.  Huang \& Gies show
that the H$\gamma$ line strength and equivalent width can be used as
starting parameters in a line profile fit to obtain unique values of
$T_{\rm eff}$, $\log g$, and their corresponding errors.  We used their
procedure to measure these quantities from our observed H$\gamma$ line
profiles.

Among the hotter B-type stars, non-LTE effects alter the equivalent
width of the H$\gamma$ line, and thus the LTE Kurucz model line profiles
systematically underestimate $T_{\rm eff}$.  Therefore we used the new
TLUSTY BSTAR2006 grid of metal line-blanketed, non-LTE, plane-parallel,
hydrostatic model spectra \citep{lanz2007} to measure $T_{\rm eff}$ and
$\log g$ for those stars with $T_{\rm eff} > 15,000$ K.  We used their
models with solar metallicity and helium abundance and a microturbulent
velocity of 2 km~s$^{-1}$.  The grid includes $T_{\rm eff}$ from
15,000--30,000 K in increments of 1,000 K and $\log g$ from 1.75--4.75
in increments of 0.25 dex.  For these hot stars we measured $T_{\rm
eff}$ and $\log g$ by comparing the H$\gamma$ line profile to the
rotationally and instrumentally broadened model spectral line profiles
at each value in the grid, minimizing rms$^2$ across the line region. 
We then refined our measurements to a higher precision using a linear
interpolation between the available line profiles in the grid.  Finally,
we determined the errors, $\Delta T_{\rm eff}$ and $\Delta \log g$, from
the values which produce a $\rm rms^2$ no more than $2.7 \, \rm
rms^2$/$N$ greater than the minimum rms$^2$.  Our measurements of
$T_{\rm eff}$ and $\log g$, with their corresponding errors, are listed
in columns 5--8 of Table \ref{params}.

Many of the B stars in our sample are rapidly rotating with $V \sin i >
200$ km~s$^{-1}$.  Such a rapidly rotating star will experience strong
centrifugal forces that distort the star into an oblate spheroidal
shape, as recently found for the star Regulus \citep{mcalister2005}. 
The surface gravity at the equator can therefore be much lower than at
the poles.  For such a rapid rotator, our measured $T_{\rm eff}$ and
$\log g$ represent the average across the visible stellar hemisphere and
are therefore biased toward lower values, causing the star to appear
more evolved.  Because the polar regions are not distorted, the surface
gravity at the poles is a better indicator of the evolutionary state of
the star.  \citet{huang2006b} performed detailed spectroscopic modeling
of such distorted rotating stars to determine a statistical correction
factor for $\log g$, averaged over all possible $i$, for a variety of
stellar models.  We made a bilinear interpolation between their models
to convert our measured $\log g$ to $\log g_{\rm polar}$ for a more
accurate comparison between slow and rapid rotators. 

For each B star, we also measured its mass, $M_\star$, and radius,
$R_\star$, by interpolating between the evolutionary tracks for
non-rotating stars from \citet{schaller1992}.  The errors $\Delta
M_\star$ and $\Delta R_\star$ correspond to our measured $\Delta T_{\rm
eff}$ and $\Delta \log g$.  Our results for $\log g_{\rm polar}$,
$M_\star$, $\Delta M_\star$, $R_\star$, and $\Delta R_\star$ are also
listed in Table \ref{params}, columns 9--13.

\subsection{$T_{\rm eff}$ and $\log g$ Measurements of Be stars}

Star No.\ 196 did not show any sign of Be emission until our most recent
observations, hence we included it among the normal B stars and measured
$T_{\rm eff}$ and $\log g$ from the H$\gamma$ line in its 2006 blue
spectrum.  For other Be stars in our blue spectra, the above method to
measure $T_{\rm eff}$ and $\log g$ was not useful because the equivalent
width of the H$\gamma$ line may be decreased by emission in the line,
even if the line profile does not exhibit obvious signs of emission. 

Our first attempt to measure the Be stars' $T_{\rm eff}$ and $\log g$
relied upon the \ion{He}{1} $\lambda\lambda$ 4143, 4388, 4471 lines.  We
fit these three lines using TLUSTY model spectra using the same
procedure described above for the H$\gamma$ line.  However, these
\ion{He}{1} line strengths are less sensitive to $T_{\rm eff}$, and the
line wings show only a very small dependence on $\log g$.  Therefore
these line fits resulted in very large errors for both parameters in
many cases.  Furthermore, the Be disks contribute continuum flux that
dilutes the apparent strength of the \ion{He}{1} lines, so $\log g$
values measured from these lines do not always agree well.

\begin{figure}
\includegraphics[angle=90,scale=0.3]{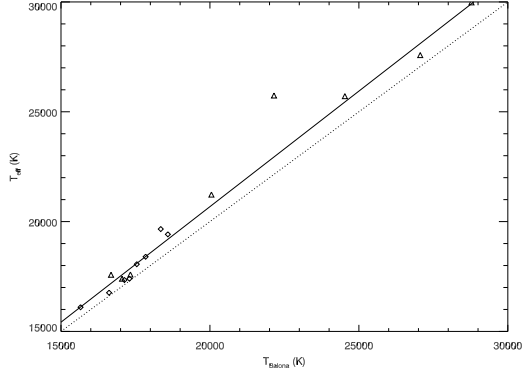}
\caption{$T_{\rm eff}$ measured for the B-type temperature calibration stars from this work (\textit{diamonds}) and from Napiwotzki (1993; \textit{triangles}) compared to their calculated $T_{\rm Balona}$ \citep{balona1984}.  A linear fit to the two temperature scales (\textit{solid line}) and the 1:1 agreement (\textit{dotted line}) are also shown.
\label{tempcalibration}
}
\end{figure}

To improve our $T_{\rm eff}$ measurements for the Be stars, we turned to
available Str\"omgren $m_1$, $c_1$, and $\beta$ indices for our targets
(\citealt{shobbrook1985, shobbrook1987}; WEBDA).  Several temperature
relations for Str\"omgren indices are available in the literature (see
\citealt{napiwotzki1993} and references therein), so we began by using
our normal B-type stars with $T_{\rm eff} > 15000$~K as calibrators to
determine the best relation for our data.  We adopted the reddening
value of $E(b-y) = 0.15$ (corresponding to $E(B-V) = 0.2$) for NGC 3766
\citep{shobbrook1985, moitinho1997}.  Eight stars in our sample (Nos.\
16, 42, 49, 54, 57, 161, 170, and 196) have available Str\"omgren
indices as well as $T_{\rm eff}$ measured from our $H\gamma$ line fits,
and Napiwotzki et al.\ provide eight calibrators with well-known $T_{\rm
eff} > 15000$~K (measured from their absolute integrated stellar flux)
and available Str\"omgren data.  For these 16 B stars, we found the best
overall agreement using the temperature relation from
\citet{balona1984}, shown in Figure \ref{tempcalibration}.  However, we
found that $T_{\rm Balona}$ systematically underestimated $T_{\rm eff}$,
and a correction factor was necessary to improve their agreement.  We
performed a linear fit to the data and found the relationship
\begin{equation}
T_{\rm eff, fit} = 1.052 \: T_{\rm Balona} - 359.636 \; \rm K
\end{equation}
(excluding one outlying point from \citealt{napiwotzki1993}).  The slope
of this correction is virtually identical to the values found by both
\citet{gies1992} and \citet{cunha1994}.  After applying this correction,
we found a mean scatter of 264 K from our two independent measurements
of $T_{\rm eff}$ for the B stars in NGC 3766.

\placefigure{tempcalibration}

\begin{figure}
\includegraphics[angle=90,scale=0.3]{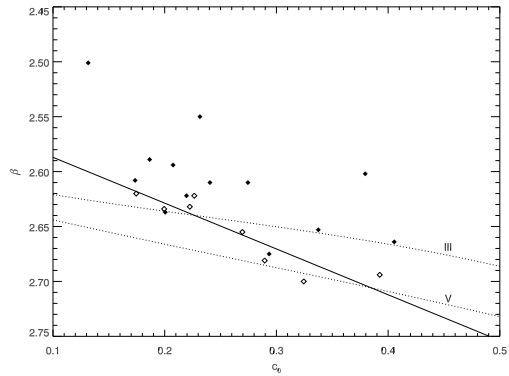}
\caption{Str\"omgren $c_0$ index plotted against the $\beta$ magnitude for the B-type temperature calibration stars from this work (\textit{diamonds}).  Be stars are also plotted (\textit{filled diamonds}) to illustrate the contamination in $\beta$ due to their disk emission.  We also show the ($c_0, \beta$) relations for luminosity class V and III stars (\citealt{balona1984b}; \textit{dotted lines}) and our linear fit to this evolving population (\textit{solid line}).
\label{c0_Hbeta}
}
\end{figure}

$T_{\rm Balona}$ relies upon the dereddened $c_0$ index as well as the
narrow-band $\beta$ magnitude, so determining an accurate temperature
for the Be stars also requires confidence in $\beta$.  However, the Be
stars' emission makes the $\beta$ magnitude highly unreliable. 
Therefore we used the B star calibrators from our sample (listed above)
to investigate several ($c_0, \beta$) relations in the literature
\citep{crawford1978, balona1984b}.  The ($c_0, \beta$) diagram is
essentially a H-R diagram that reveals temperature and evolutionary
trends in a population.  In Figure \ref{c0_Hbeta}, we show that the
values of $c_0$ and $\beta$ for these B-type calibrators generally agree
with the relations for luminosity class III and V stars from
\citet{balona1984b}.  However, the calibration stars have $3.43 < \log g
< 3.98$ ($3.49 < \log g_{\rm polar} < 4.17$) since these hot stars are
evolving along the MS, and neither ($c_0, \beta$) relation can be
applied to the entire population.  Therefore we performed a linear fit
to account for the range in evolution, and we found the relationship
\begin{equation}
\beta_{\rm fit} = 0.417 \: c_0 + 2.545
\end{equation}
among the B stars in NGC 3766 with $T_{\rm eff} > 15000$~K.  The mean
scatter between $\beta_{\rm fit}$ and the measured $\beta$ is 0.010,
which implies an additional error of 144 K in $T_{\rm eff}$ using the
corrected $T_{\rm Balona}$ relation above. 

\placefigure{c0_Hbeta}

\citet{shobbrook1985, shobbrook1987} presented Str\"omgren photometry
for nine Be stars that have accompanying blue spectra in this work
(excluding No.\ 196, which we used as a calibrator) and four Be stars
with only red spectra in this work.  Thus we measured $T_{\rm eff}$ for
all 13 of these Be stars using the adopted $\beta_{\rm fit}$ and the
corrected $T_{\rm Balona}$ as described above.  We adopt a total error
of $(264^2 + 144^2)^{0.5} = 301$~K for $T_{\rm eff}$ measured with this
method, and the results are listed in columns 5--6 of Table
\ref{params}.

Using this technique, only the ``shell'' Be star No.\ 154 results in
$T_{\rm eff} < 15000$~K .  The comparisons between $T_{\rm Balona}$ and
our independently measured $T_{\rm eff}$ are less reliable below this
temperature threshold, hence we used only B star calibrators with
$T_{\rm eff} > 15000$~K.  However, spectroscopy is even less likely to
produce accurate measurements for No.\ 154 since its spectrum is
contaminated by \ion{He}{1} emission and the metal lines are
exceptionally strong due to the disk's edge-on orientation.  Therefore
we include in Table \ref{params} its $T_{\rm eff}$ from the corrected
relations of \citet{balona1984}, but the errors are somewhat higher than
for the other Be stars.

The values of $\log g$ are more strongly dependent on $\beta$, so we
were reluctant to use our $\beta_{\rm fit}$ with the Str\"omgren
relation for $\log g$ given by \citet{balona1984}.  Instead, we used the
calculated $T_{\rm eff}$ to determine the Be stars' bolometric
corrections, BC, from \citet{lanz2007}, at first assuming $\log g=4.0$. 
For the cooler star No.\ 154, we interpolated the BC from the values for
MS stars given by \citet{cox2000}.  We calculated each stellar radius,
$R_\star$, and luminosity, $L_\star$, using the measured $T_{\rm eff}$,
BC, $V$ magnitude (\citealt{shobbrook1985, shobbrook1987}; WEBDA),
distance modulus $(V-M_V)_0 = 11.73 \pm 0.33$ \citep{moitinho1997}, and
$E(B-V) = 0.2 \pm 0.1$ \citep{shobbrook1985, moitinho1997}.  We measured
the stellar mass, $M_\star$, from the computed $T_{\rm eff}$ and
$L_\star$ by interpolating between the evolutionary tracks of
\citet{schaller1992}.  Finally, we obtained a preliminary value of $\log
g$ from $M_\star$ and $R_\star$.  Since the BC is weakly dependent on
$\log g$, we improved the BC from the initial estimate and iterated to
compute the final $\log g$.  We adopt a formal error in $\log g$
computed from $\Delta T_{\rm eff}$ and the quoted errors in $(V-M_V)_0$
and $E(B-V)$. 

We note that the scatter between the calibrators' $\log g_{\rm polar}$
and their Str\"omgren $\log g$ is identical to the formal error.  Based
on this good agreement, we do not perform any further correction to
obtain $\log g_{\rm polar}$ for the Be stars measured with this
technique.  However, we show below that the Be stars are more rapidly
rotating than the normal B stars, which may make them appear slightly
more evolved and artificially brightened.  Thus these values of $\log
g_{\rm polar}$ are lower limits.  The final parameters for these Be
stars are listed in Table \ref{params}.

\begin{figure}
\includegraphics[angle=90,scale=0.3]{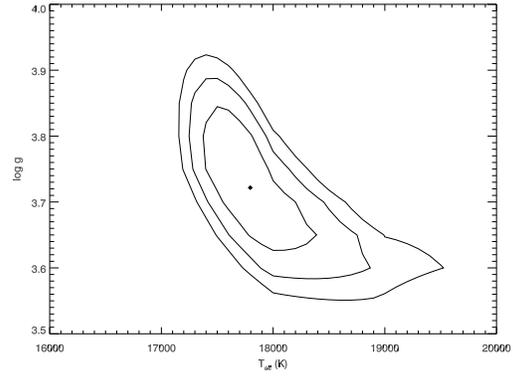}
\caption{A contour plot of the $1\sigma$, $2\sigma$, and $3\sigma$ errors in the \ion{He}{1} $\lambda4388$ line fit for No.\ 31.  We adopted the center of the $1\sigma$ error region as the best fit value, and the extent of this region indicates the sizes of the error bars for $T_{\rm eff}$ and $\log g$. 
\label{contour}
}
\end{figure}

Two Be stars in our sample of blue spectra have no available Str\"omgren
data in the literature.  By coincidence, these stars (Nos.\ 31 and 127)
have two of the best $T_{\rm eff}$ and $\log g$ measurements from our
preliminary \ion{He}{1} line fits.  Therefore we adopt the mean
measurements from the \ion{He}{1} $\lambda\lambda$ 4143, 4388, 4471
lines, with a formal error determined from the values which produce a
$\rm rms^2$ no more than $2.7 \, \rm rms^2$/$N$ greater than the minimum
rms$^2$.  A contour plot of the errors in $T_{\rm eff}$ and $\log g$ for
No.\ 31 is shown in Figure \ref{contour}, and the final values are
listed in Table \ref{params}.  We determined their $\log g_{\rm polar}$
using the same method as the B stars.

\placefigure{contour}

\subsection{Discussion of Physical Parameters}

\begin{figure}
\includegraphics[angle=90,scale=0.3]{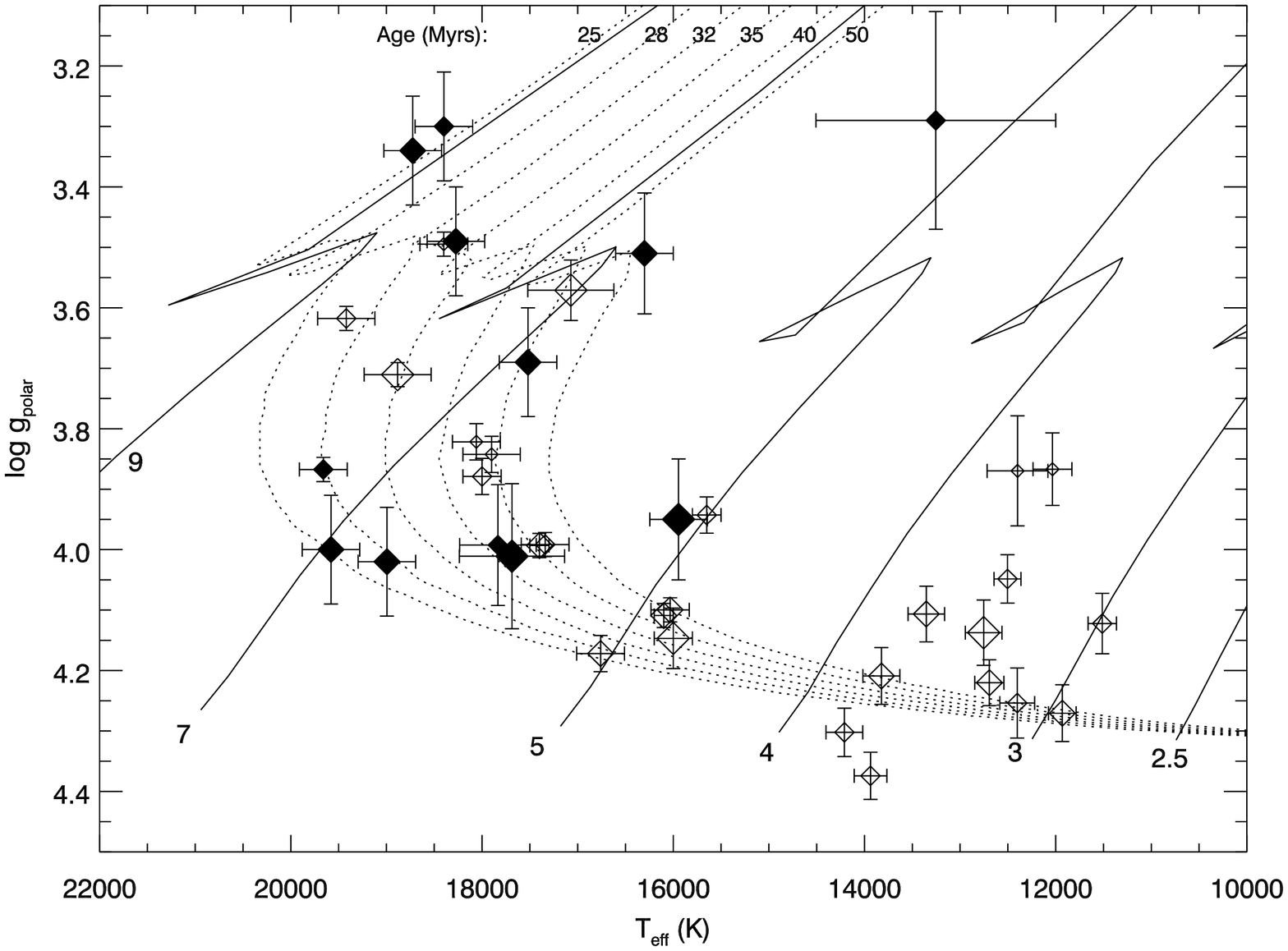}
\caption{$T_{\rm eff}$ and $\log g_{\rm polar}$ are plotted with the evolutionary tracks of \citet{schaller1992} (\textit{solid lines}) and isochrones of \citet{lejeune2001} (\textit{dotted lines}).  The zero-age MS mass of each evolutionary track is labeled along the bottom, and the age of each isochrone is labeled along the top.  Normal B-type stars are shown as open diamonds while Be stars are filled diamonds, and each symbol size is proportional to the star's $V \sin i$.  \label{tlogg}
}
\end{figure}

The resulting values of $T_{\rm eff}$ and $\log g_{\rm polar}$ are
plotted in Figure \ref{tlogg} with the corresponding evolutionary tracks
for non-rotating stars from \citet{schaller1992}.  The symbol sizes are
proportional to $V \sin i$ to investigate the relation between rotation
and the apparent evolutionary state of each star, but no trends are
observed.  Also in Figure \ref{tlogg}, we plot the isochrones for 25--50
Myr populations from \citet{lejeune2001}.  Our distribution of $T_{\rm
eff}$ and $\log g_{\rm polar}$ are generally consistent with a
population within this range, indicating an age slightly greater than
previous estimates for NGC 3766. 

The masses of the rapidly rotating stars may be overestimated since the
evolutionary tracks do not account for rotation.  We minimize this
effect by using the corrected $\log g_{\rm polar}$ for all stars, but
slight mass differences may still be present since rapid rotation is
expected to alter the evolution \citep{heger2000, meynet2000}. 
\citet{mcalister2005} found a small, 15\% mass discrepancy for the rapid
rotator Regulus when comparing its mass derived from non-rotating
evolutionary tracks and its true mass from a detailed spectroscopic and
interferometric analysis. 

We find that the Be stars are generally among the hotter stars in our
sample, consistent with our earlier findings from a photometric
investigation of 48 open clusters that Be stars are preferentially found
among the more luminous cluster members \citep{mcswain2005b}.  There are
no systematic differences between the $\log g_{\rm polar}$ of the Be
star and normal B star populations of the cluster, indicating that the
Be stars in NGC 3766 are distributed across a range of $\log g_{\rm
polar}$ and are not associated with any particular stage of the MS
evolution of B-type stars.  This is also consistent with our earlier
results \citep{mcswain2005b} and with \citet{zorec2005}, who performed
an evolutionary study of field Be stars.  On the other hand,
\citet{levenhagen2006} found that field Be stars are preferentially
found at later stages of the MS evolution.  They did not perform any
corrections for gravity darkening among these rapid rotators, so their
Be stars may be found closer to the zero-age MS than their results
suggest. 

\placetable{params}
\placefigure{tlogg}

%%%%%%%%%%%%%%%%%%%%%%%%%%%%%%%%%%%%%%%%

\section{Rotational Velocities}

Be stars are often described as a population of rapidly rotating B-type
stars with a true rotational velocity $V_{\rm rot}$ comparable to the
critical velocity $V_{\rm crit}$.  (Note that $V_{\rm rot}$ should not
be confused with $V$ in our measured $V \sin i$, which may be subject to
systematic errors such as gravitational darkening or weak emission in
the \ion{He}{1} lines.)  However, a recent study by \citet{cranmer2005}
has cast some doubt on their fast rotation.  He compared available $V
\sin i$ measurements in the \citet{yudin2001} database of Oe, Be, and Ae
stars to the predicted distribution of $V \sin i$ accounting for gravity
darkening, limb darkening, and observational effects.  He found that the
Be stars of the Yudin database have intrinsic rotations between
40--100\% critical, with more early-type Be stars having significantly
subcritical rotation.  However, the Yudin database is a compilation of
measurements from many different authors and instrumental setups, and
therefore the available $V \sin i$ may contain significant systematic
differences.  Our measurements of $V \sin i$ rely upon data of identical
origin and measurements of the same spectral lines in nearly every case. 
Therefore we have greatly reduced the systematic differences among our
Be star measurements, and our measurements are also a reliable
comparison of Be stars relative to normal B stars.

A rotationally distorted star has an equatorial radius $R_{\rm e} = 1.5
R_{\rm p}$ in the Roche approximation; here, $R_{\rm p}$ is the star's
polar radius.  For simplicity, we assume $R_{\rm p}$ is equal to the
radius $R_\star$ derived from the position in Figure \ref{tlogg} and
given in Table \ref{params}.  The resulting critical velocity is
\begin{equation}
V_{\rm crit} = \sqrt{\frac{G M_{\star}}{R_{\rm e}}}
\end{equation}
and is included in Table \ref{params}.  A small mass discrepancy, as
found for Regulus \citep{mcalister2005}, will not affect $V_{\rm crit}$
significantly since a 15\% error in mass produces only a 4\% error in
$V_{\rm crit}$.  Here we investigate the rotational properties of
several edge-on Be stars in our sample and compare the distribution of
Be star velocities with the normal B stars to compare the two
populations.

Two Be stars in our sample, Nos.\ 92 and 139, show H$\alpha$ emission
only in the line wings, with a deep absorption profile, suggesting the
disks are observed nearly edge-on.  This is probably a good assumption
for No.\ 139 since it has $V/V_{\rm crit} \geq 0.7$.  However, No.\ 92
may not be edge on since its H$\alpha$ profiles between 1985$-$1990
\citep{balona1991} do not resemble the profiles we observe.  Either the
disk is precessing or other structural changes have occurred.  A third
star, No.\ 154, exhibits a shell spectrum with strong metal lines that
also suggests an edge-on orientation. 

If we assume that all of the Be stars have $V/V_{\rm crit} = 0.95$, then
our measured $V \sin i$ indicate that stars 92, 139, and 154 have $i =
34^\circ$, $48^\circ$, and $15^\circ$, respectively.  Assuming a slower,
$V/V_{\rm crit} = 0.70$ increases the derived values to $i = 49^\circ$,
$90^\circ$, and $20^\circ$ respectively, but the low $i$ for Nos.\ 92
and 154 are still inconsistent with the observed line profiles. 
However, if we fix $i = 80^\circ$ for all three stars, we find that
$V/V_{\rm crit} = 0.53$ for No.\ 92, $V/V_{\rm crit} = 0.71$ for No.\
139, and $V/V_{\rm crit} = 0.24$ for No.\ 154.  \citet{townsend2004}
show that measured $V \sin i$ may be too low due to gravitational
darkening in highly distorted, rapidly rotating stars.  But even
accounting for a 20--33\% underestimate in $V \sin i$ for Nos.\ 92 and
154, we do not find that these two Be stars are near critical rotation. 

\begin{figure}
\includegraphics[angle=90,scale=0.3]{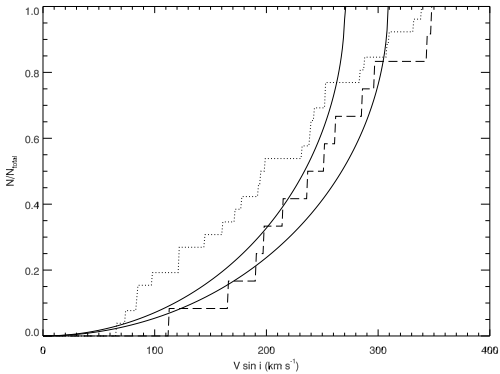}
\caption{Cumulative distribution function of $V \sin i$ for the Be stars (\textit{dashed line}) and the normal B-type stars (\textit{dotted line}).  Theoretical distributions of rapidly rotating stars with random orientation and $V = 0.7 \; V_{\rm crit}$ and $0.8 \; V_{\rm crit}$ are also shown (\textit{solid lines}; assumes the mean value for the Be stars, $V_{\rm crit} = 386$~km~s$^{-1}$).
\label{vsini}
}
\end{figure}

The distributions of $V \sin i$ for both the Be stars and normal B stars
are ploted in Figure \ref{vsini}.  Although this sample of 38 stars is
small, this study is the first to measure $V \sin i$ for both groups
using a consistent method and a uniform data set.  Our measured $V \sin
i$ for the Be stars may be only lower limits, but we clearly find that
the Be stars of NGC 3766 are more rapidly rotating than the normal B
star population.  In Figure \ref{vsini}, we also include theoretical
distributions of two uniform, rapidly rotating populations with $V = 0.7
\; V_{\rm crit}$ and $V = 0.8 \; V_{\rm crit}$, assuming $V_{\rm crit} =
386$~km~s$^{-1}$ (the mean value among the Be stars, excluding No.\ 154
due to its large measurement errors).  The distribution of Be stars is
consistent with $0.7 \; V_{\rm crit} < V < 0.8 \; V_{\rm crit}$ in most
cases.  However, if each measured $V \sin i$ is underestimated by
20--33\% due to gravitational darkening, as claimed by
\citet{townsend2004}, the distribution of Be star velocities is
consistent with $V_{\rm rot} \geq 0.84 \; V_{\rm crit}$.  Nos.\ 92 and
154 represent significant exceptions to this rule, as discussed above. 

\placefigure{vsini}

%%%%%%%%%%%%%%%%%%%%%%%%%%%%%%%%%%%%%%%%%%%

\section{Other Results from Spectra}

We do not wish to broaden the scope of this paper to discuss the radial
velocities and helium abundances observed in our spectra.  However, we
found several instances that suggest these topics are worth exploring in
future studies of NGC 3766. 

Figure \ref{specvar1} includes the variable H$\alpha$ profile of No.\
61.  The absorption line is deep and narrow in 2005, yet wide and
shallow in 2006 and 2007.  These line profile changes are not consistent
with a Be star, and they probably indicate line blending in a
double-lined spectroscopic binary (SB2).  Likewise, No.\ 197 may be an
SB2.  We observed double \ion{Mg}{2} 4481 lines in 2006, and Figure 3
shows a large wavelength shift in its H$\alpha$ line profile between
2005$-$2006.  There is possible line blending visible in the 2005 and
2007 H$\alpha$ profiles. 

We measured the physical parameters of the B stars using spectral models
that have solar helium abundances, and most were consistent with a solar
abundance.  However, we noticed a large number that have abnormal He
abundances; Nos.\ 41, 55, 94, 118, and 178 appear to be He strong, while
Nos.\ 126, 129, 170, 173, and 197 appear He weak.  Since all cluster
members are expected to have the same He abundance, this may indicate
widespread magnetic fields in the cluster which can produce nonuniform
distributions of He across the stellar photospheres (see discussion of
\citealt{huang2006b}).  The He weak stars may also be binaries with
cooler companions adding flux but diluting the appearance of the
\ion{He}{1} lines.  

Finally, we used our spectra to reinvestigate the reddening and distance
to NGC 3766.  For the nonemission B stars with $\log g_{\rm polar} >
3.9$, we used the MS relation of \citet{harmanec1988} to assign a
spectral type to each star based on its measured $T_{\rm eff}$.  We then
used the MS relation of \citet{wegner1994} to find the stars' intrinsic
colors, $(B-V)_0$.  Observed $B-V$ colors were obtained from WEBDA.  For
the 17 B-type, MS stars in our sample, we found $E(B-V) = 0.223 \pm
0.030$.  We then used the B stars' $V$ magnitude (WEBDA), BC
(\citealt{lanz2007}, \citealt{malagnini1986}), and our measured
$R_\star$ to compute the distance modulus, $(V-M_{\rm V})_0$.  The
resulting $(V-M_{\rm V})_0 = 11.42 \pm 0.15$.  To determine the mean
$(V-M_{\rm V})_0$, we excluded No.\ 45 with a somewhat lower $(V-M_{\rm
V})_0$ of 10.3, probably a foreground star. 

While the B stars of NGC 3766 are rotating more slowly than the Be
stars, they have a mean $V \sin i/V_{\rm crit} = 0.5$.  Assuming a
random distribution of $i$, this corresponds to a mean $V/V_{\rm crit} =
0.7$.  Such rapidly rotating stars are probably rotationally distorted
with surface areas $\approx 1.2$ times larger than spherical star with
$R_\star = R_{\rm polar}$, causing them to appear overluminous.  The
distance modulus is then underestimated by a factor of $2.5 \log 1.2 =
0.21$.  With a realistic distribution of $V$ and $i$, this correction
may not be appropriate for all stars, and the range in $\Delta (V-M_{\rm
V})_0$ is $0.02-0.30$ for $0.28 \leq V/V_{\rm crit} \leq 0.83$. 
We apply this correction to find a final $(V-M_{\rm V})_0 = 11.6 \pm
0.2$ for NGC 3766, corresponding to a distance of $1.9-2.3$ kpc.  Our
new values of $E(B-V)$ and $(V-M_{\rm V})_0$ are highly consistent with
previous results for this cluster.

%%%%%%%%%%%%%%%%%%%%%%%%%%%%%%%%%%%%%%%%%

\section{Be Star Variability \label{var}}

While the long term H$\alpha$ variability of Be stars has often been
noted in the literature (e.g.\ \citealt{porter2003}), few studies have
attempted long term monitoring of a large sample of Be stars to quantify
their variability.  \citet{hubert1998} investigated 273 bright Be stars
with $V < 7.5$ that were observed by Hipparcos between 1989 August and
1993 August.  They found that early type Be stars exhibit a very high
degree of variability while most late Be stars maintained a constant
magnitude during the duration of the Hipparcos mission.  They also
identified 14 Be stars with recurrent short-lived outbursts ($0.06 \leq
\Delta Hp \leq 0.3$ over timescales of 50$-$500 days) and 8 with
long-lived outbursts ($\Delta Hp \geq 0.12$ over timescales of $> 500$
days) in the sample.  However, the Hipparcos magnitude, $Hp$, covers a
broad waveband ranging from 3400 to 8500 \AA, and it traces large scale
variations in the continuum from scattered light associated with the
disk.  H$\alpha$ spectroscopy is better suited to measuring variations
over a larger dynamic range in the disks of Be stars of all spectral
types. 

Figures \ref{specvar1}$-$\ref{specvar3} show the H$\alpha$ line profile
variations of all 47 stars in our spectroscopic sample.  We were
intrigued to discover a new Be star (No.\ 130) in NGC 3766 based on our
comparison of its H$\alpha$ line profiles.  In 2005 February, the
absorption line was broad and shallow, with some bumps that suggest a
weak emission disk, nonradial pulsations, or simply noise.  By 2006 May,
however, the depth of the line had increased while its width remained
nearly the same, but a slight asymmetry in the line suggested that
perhaps a very weak disk may still have been present.  The line remained
in absorption during our most recent 2007 observations.  While an
inverse relation between the line width and depth is associated with
line blending in an SB2, the observed variations in No.\ 130 cannot be
attributed to such line blending.  Instead, the three sets of
observations indicate the presence of a weak circumstellar disk in 2005
that largely disappeared by 2006.

In fact, we were startled to find that a total of 11 Be stars in our
sample exhibit significant changes in their disk state between
2003--2007.  Stars 25, 31, 73, 83, 92, 98, 119, 130, 133, 139, and 196
each show H$\alpha$ in absorption in at least one of the four epochs
available, yet each star also experiences at least one epoch when the
H$\alpha$ line is partially or fully filled with emission.  These
observations suggest that the mechanism responsible for the disk
formation is unstable over timescales of only a few years, consistent
with the results of \citet{hubert1998}.  The disk growth of No.\ 31
during 2007 reveals that the disk formation can be very rapid, only
requiring a few days or weeks, as observed in other Be stars (e.g.
\citealt{grundstrom2007}).  However, the high fraction of Be stars that
show significant variability in a single cluster is unprecedented, and
the cluster should be monitored with increased frequency to measure an
accurate timescale of the disk state changes.

Five additional Be stars (47, 127, 154, 198, and 200) show H$\alpha$
emission in every observation, although the emission strength is usually
variable.  Such variations in H$\alpha$ emission strength are typical
among Be stars. 

To quantify the observed changes in the Be star disks, we measured the
equivalent width, $W_\lambda$, of the H$\alpha$ line in each of our red
spectra by normalizing each spectrum to a unit continuum and integrating
over the line profile.  Columns 3$-$6 of Table \ref{eqwidth} give the
measured values of $W_{\lambda}$ for each spectrum.  The error in each
measurement is about 10\% due to noise in the continuum region.

With multiple epochs of H$\alpha$ observations available for so many
variable Be stars, it is worthwhile to estimate the changing size of the
circumstellar disks.  \citet{grundstrom2006} describe a simple model to
measure the ratio of the projected effective disk radius to the stellar
radius, $R_{\rm disk}$/$R_{\star}$, and the density at the base of the
disk, $\rho_0$, using $W_{\lambda}$ and $T_{\rm eff}$ as input
parameters.  We determined the disk inclination $i$ by assuming that
each star is rotating with $V$ at 70\% of $V_{\rm crit}$ in most cases. 
We also assumed a disk truncation radius of 100 $R_{\star}$, the nominal
value unless a close binary companion is present.  The disk temperature
is assumed to be constant at $0.6 \; T_{\rm eff}$. 

To estimate the total masses of the disks, we used an axisymmetric,
isothermal density distribution,
\begin{equation} \label{densityeqn}
\rho(r,z) = \rho_0 \left (\frac{R_\star}{r} \right )^n \exp \left[-\frac{1}{2} 
\left(\frac{z}{H(r)}\right)^2 \right]
\end{equation}
\citep{carciofi2006} and a radial density exponent $n = 3$, typical of
other Be star disks \citep{gies2007}.  The scale height of the disk is
\begin{equation}
H(r) = H_0 \left( \frac{r}{R_{\star}}\right)^\beta,
\end{equation}
where
\begin{equation}
H_0 = \frac{a}{V_{\rm crit}} R_{\star},
\end{equation}
\begin{equation}
a = \sqrt{\frac{kT}{\mu m_{\rm H}}},
\end{equation}
and $\beta = 1.5$ for an isothermal disk \citep{bjorkman2005, carciofi2006}.  

The resulting disk properties are listed in Table \ref{bevar} for each
Be star with an available $T_{\rm eff}$ measurement.  The time of each
observation is listed in column 2, and the dates for the CTIO 4m+Hydra
observations are less precise since these $W_{\rm H\alpha}$ were
measured from co-added spectra rather than individual spectra obtained
with the CTIO 1.5m telescope.  Our determined values for $i$, $\rho_0$,
$R_{\rm disk}/R_{\star}$, and $M_{\rm disk}$ are listed in columns 4--7. 

Certainly, these disk measurements should be viewed with caution since
the true density profiles may be quite different from the assumed
distribution. 
The disk density exponent may be different from our assumed value of
$n=3$, and the resulting mass estimates may differ from our results by
orders of magnitude. Contemporaneous optical and infrared spectra have
also revealed evidence of density waves that alter the azimuthal disk
structure \citep{wisniewski2007}.  Finally, the disks are likely not
isothermal as we assumed, and the true thermal structure may be very
complex \citep{carciofi2006}.  Therefore the processes associated with
the H$\alpha$ emission profile are much more complex than assumed in the
model of \citet{grundstrom2006}, but by using their simple model we
obtain relative estimates of the variations in disk base density that
are sensible provided the disk density exponent is assumed constant in
time.

\placetable{bevar}

%%%%%%%%%%%%%%%%%%%%%%%%%%%%%%%%%%%%%%%%%%

\section{Disk Formation by Nonradial Pulsations}

The estimates for $\rho_0$ and $M_{\rm disk}$ above are not sensitive to
$i$ and offer a consistent method to measure changes in the size of the
disks.  Thus we can estimate the disk growth rate (or dissipation rate),
$\Delta M_{\rm disk}/\Delta t$.  Our generally sparse observations
cannot determine whether $\Delta M_{\rm disk}/\Delta t$ is constant or
highly variable over timescales less than one year; however there is an
indication from the two closely spaced observations of Nos.\ 25 and 31
that the disk size can change rapidly.  We provide our measured $\Delta
M_{\rm disk}/\Delta t$ between each pair of observations in Table
\ref{bevar}.  The long-term buildup rates are comparable for every Be
star in our sample, hinting that all of the disks might be formed by the
same mechanism.  Finally, the buildup and dissipation rates are also
comparable in magnitude. 

Many Be stars are known to exhibit nonradial pulsations (NRP) with the
mode $\ell = 2$, $m = \pm2$ \citep{rivinius2003}, and such pulsations
are commonly proposed as the source of kinetic energy to inject material
into the disk \citep{porter2003}.  The $m = +2$ pulsational mode is
retrograde, and naively it will counteract the rotational velocity of
the star and hinder disk formation.  However, \citet{townsend2005} shows
that retrograde mixed modes behave in a way that could contribute to Be
disk formation.  Although their phase velocity is retrograde, their
group velocity is prograde.  The density enhancements occur when the
pulsational velocity perturbation is in the same direction as the
rotation -- a configuration favorable to mass ejection.  He also found
the mixed mode instability strip likely overlaps with the temperatures
and spectral types of known Be stars if their rotation is nearly
critical. 

Observations of known $\beta$ Cephei pulsators and slowly pulsating B
stars indicate that their pulsation modes have velocity amplitudes on
the order of 10 km~s$^{-1}$ (combining the radial, azimuthal, and
longitudinal components of the total velocity vector;
\citealt{deridder2001}).  \citet{owocki2005} shows that the velocity
needed for ejection into the disk is $\Delta V_{\rm orb} = V_{\rm crit}
- V_{\rm rot}$.  For a weak atmospheric process such as NRP, the
atmospheric sound speed $a \sim 12$ km~s$^{-1}$ must be comparable to
$\Delta V_{\rm orb}$.  If we assume that the stars have $V_{\rm rot} =
0.95 \; V_{\rm crit}$, then $\Delta V_{\rm orb} \sim 20$ ~km~s$^{-1}$,
highly comparable to $a$ and the observed pulsation velocities in other
NRP stars \citep{deridder2001}.  As we show above, this assumption of
near critical rotation is reasonable for most of the Be stars in NGC
3766.

To compare the NRP velocity with the outflow velocity required to fill
the Be disks, we use the equation of mass continuity for an equatorial
outflow with velocity $v_{\rm r, eq}$:
\begin{equation}
\sigma (r) = \frac{\Delta M_{\rm disk}/\Delta t}{2 \pi r \: v_{\rm r, eq}}
\end{equation}
where the surface density of the disk is 
\begin{equation}
\sigma (r) = \int^{\infty}_{-\infty} \rho (r,z) \: dz.
\end{equation}
Since we remove the vertical dependence of the disk density by using its
surface density, this method effectively assumes that all of the
material leaves the star through a cylindrical surface of arbitrary
height.  For a disk reaching down to the stellar surface, the mass must
flow across this surface, so we use the surface density at the stellar
surface, $\sigma(R_{\rm eq})$, in the continuity relation. Since the
$\ell = 2$, $m = \pm 2$ modes have the greatest pulsational amplitude at
the stellar equator, where the effective surface gravity is also a
minimum, it is reasonable to expect the majority of the disk material to
be ejected from the equator in the NRP model.  The observed disk
formation rates are generally $\Delta M_{\rm disk}/\Delta t = 10^{-12} -
10^{-11}$ ~$M_\odot$ ~yr$^{-1}$ with $\rho_0 = 10^{-12}$ ~g~cm$^{-3}$
from Table \ref{bevar}.  This implies a very slow surface flow on
average, $v_{\rm r, eq} \sim .01 - 0.1$ ~km~s$^{-1}$.  However, a few
observations reveal that the surface flow can be an order of magnitude
faster, but this is still consistent with NRP. 

Likewise, we use our observed disk densities to demonstrate that NRP are
a capable source of angular momentum, $L$, for disks.  The rotational
inertia, $I$, of a thin disk shell is given by
\begin{equation}
I(r) = 2 \pi r^3 \sigma(r) \: dr
\end{equation}
and the total angular momentum of the disk, extending from the stellar
surface to $100 R_\star$, is
\begin{equation}
L = \int^{100 R_\star}_{R_{\rm eq}}  I(r) \: \omega(r) \: dr.
\end{equation}
where $\omega(r) = (GM_\star)^{0.5} \: r^{-1.5}$ is the angular velocity
of a Keplerian disk. We find that $L \sim
10^{43}-10^{44}$~g~cm$^2$~s$^{-1}$ for the observed disks, with $\Delta
L/\Delta t \sim 10^{32}-10^{36}$~g~cm$^2$~s$^{-2}$.  \citet{osaki1986}
shows that the angular momentum flux in the equatorial plane due to NRP
with $\ell = 2$, $m = \pm2$ is given by
\begin{equation}
\frac{dL}{dt} = 2 \pi R_{\rm eq}^2 \: \sigma(R_{\rm eq}) \: A^2 k \: \sin \delta.
\end{equation}
Here, $A$ and $2Ak$ are the radial and azimuthal amplitudes of
pulsation, respectively, and $\delta$ is the phase shift between the two
velocity components.  Thus we find that a small azimuthal pulsation with
$2Ak \sim 0.01 - 0.1$~km~s$^{-1}$ can, in principle, provide enough
angular momentum to eject material into the observed disks, and both the
radial and azimuthal amplitudes are highly consistent with the observed
pulsation amplitudes of other B stars with NRP \citep{deridder2001}.

Our results suggest that the NRP may be a transitory phenomenon for the
Be stars, and the disks may fill substantially during short periods of
surface activity.  Evidence for changing pulsational modes among several
stars in NGC 3766 has been observed by \citet{balona1991},
\citet{vanvuuren1988}, and \citet{balona1986}.  They found periodic
light curve variations in several Be stars in NGC 3766 (our Nos.\ 47,
92, 133, 154, and 200).  Additional Be stars, our Nos.\ 98, 130, 198
also exhibit light curve variations that are possibly periodic.  In
several cases, the shapes and possibly periods of the light curves
change with timescales of only a few weeks.  \citet{balona1991} argue
that this is due to magnetic fields, but the light curves are not
qualitatively similar to variations observed in other magnetic stars,
such as the prototype $\sigma$ Ori E \citep{oksala2007}.  The changing
light curves among NGC 3766 members are probably more consistent with
changes in active pulsation modes over short timescales.

%%%%%%%%%%%%%%%%%%%%%%%%%%%%%%%%%%%%%%%%

\section{Conclusions}

Our spectroscopic analysis of NGC 3766 has revealed that Be stars may be
much more common than we originally thought.  In our photometric study
of NGC 3766 \citep{mcswain2005b}, we found up to 13 Be stars (5
definite, 8 uncertain) out of an expected 191 B-type stars, not counting
the one Be star that saturated our photometry.  The new total of 16 Be
stars is 23\% greater.  Among these 16 Be stars, 2--5 of them appear to
have almost no disk at any given time, and an additional 2--4 have
extremely subtle emission in their H$\alpha$ line profile that could
easily be mistaken for other phenomena (such as NRP manifesting
themselves as bumps moving across the line or SB2 line blending). 
Therefore 25--50\% of the Be stars may go undetected in a single
spectroscopic observation, and photometric snapshots are even less
likely to discern such weak emitters.  We note four stars (Nos.\ 27, 45,
49, and 77) that were found to be possible or likely Be stars in the
photometric study by \citet{shobbrook1985, shobbrook1987}, but they
never showed emission during our observations and thus remain
unconfirmed.  The existence of transitory, weak disks (especially Nos.\
130 and 196) could mean that many more Be stars are waiting to be
discovered. 

For our total sample of 48 Southern open clusters in our photometric
survey, we found a low Be fraction of $2-7$\% \citep{mcswain2005b}. 
Considering the very weak disks that are observed in NGC 3766 and the
exceptionally high variability among the cluster's Be population, the
total fraction of Be stars could be much greater.  We are currently
performing a similar spectroscopic study of several other clusters from
our survey, and we will address those results in a future paper.

While the Be stars of NGC 3766 are not distinguishable from normal
B-type stars by their evolutionary states, they do form a population of
rapidly rotating stars.  With two exceptions, their measured velocities
are consistent with a uniform population of rapid rotators having $V =
0.7-0.8 \; V_{\rm crit}$.  Gravitational darkening and weak emission in
the \ion{He}{1} lines may mean that these velocities are underestimated
by as much as 33\% \citep{townsend2004}, so the true $V_{\rm rot}$ is
probably at least $0.84 \; V_{\rm crit}$.  From the measured changes in
the disks' masses and angular momenta, NRP are a capable source for the
mass flow into the equatorial plane.  The pulsations may be a transitory
phenomenon, however, and the variable nature of the Be stars probably
reflects dramatic changes in the surface activity.

%%%%%%%%%%%%%%%%%%%%%%%%%%%%%%%%%%%%%%%%%%%%%

\acknowledgments 
We thank the referee, Phil Massey, for providing comments that improved
this work.  Also, we are grateful to Giovanni Carraro, Mark
Pinsonneault, and Swetlana Hubrig for helpful discussions that
contributed to this work.  We are grateful to Yale University and the
SMARTS Consortium for providing observing time at the CTIO 1.5m
telescope.  This research has made use of the WEBDA database, operated
at the Institute for Astronomy of the University of Vienna.  MVM was
supported by an NSF Astronomy and Astrophysics Postdoctoral Fellowship
under award AST-0401460.  This work was also supported by the National
Science Foundation under grant AST-0606861 (DRG) and by NASA under grant
LTSA/NNG05GC36G (RHDT).

Facilities: \facility{CTIO}.

%%%%%%%%%%%%%%%%%%%%%%%%%%%%%%%%%%%%%%%%%%%%%

\clearpage

\clearpage

\clearpage

\clearpage

\clearpage

% Summary of observations

\begin{deluxetable}{lccclcccc}
\rotate
\tablewidth{0pt}
\tabletypesize{\scriptsize}
\tablecaption{Journal of Spectroscopy\label{spectroscopy}}
\tablehead{
\colhead{UT} &
\colhead{Range} &
\colhead{Resolving Power} &
\colhead{Number of} &
\colhead{Telescope +} &
\colhead{Slit Plate} &
\colhead{ } &
\colhead{ } &
\colhead{ } \\
\colhead{Dates} &
\colhead{(\AA)} &
\colhead{($\lambda/\Delta\lambda$)} &
\colhead{Spectra} &
\colhead{Spectrograph} &
\colhead{($\mu$m)} &
\colhead{Grating} &
\colhead{Filter} &
\colhead{Detector} }
\startdata
2003 Mar $21-22$  &  $5490-6790$  &  1800   &     20  &  CTIO 1.5m + Cassegrain  &  \nodata  &  47/1     &  GG495   &  Loral 1K$\times$1K  \\
2005 Feb 2        &  $4100-6900$  &  1900   &     47  &  CTIO Blanco 4m + Hydra  &  200      &  KPGL3/1  &  \nodata &  SITe 4K$\times$2K \\
2006 May 13       &  $3790-4708$  &  3170   &     38  &  CTIO Blanco 4m + Hydra  &  \nodata  &  KPGLD/2  &  BG39    &  SITe 4K$\times$2K \\
2006 May $14-15$  &  $5125-8000$  &  1560   &     47  &  CTIO Blanco 4m + Hydra  &  \nodata  &  KPGL3/1  &  \nodata &  SITe 4K$\times$2K \\
2007 May $4-5$    &  $5125-8000$  &  2000   &     45  &  CTIO Blanco 4m + Hydra  &  200      &  KPGL3/1  &  \nodata &  SITe 4K$\times$2K \\
2007 Jan 20       &  $5650-6790$  &  1700   &     10  &  CTIO 1.5m + Cassegrain  &  \nodata  &  47/1     &  GG495   &  Loral 1K$\times$1K \\
2007 Jan 29       &  $5650-6790$  &  1700   & \phn 2  &  CTIO 1.5m + Cassegrain  &  \nodata  &  47/1     &  GG495   &  Loral 1K$\times$1K \\
2007 Feb 2        &  $5650-6790$  &  1700   & \phn 8  &  CTIO 1.5m + Cassegrain  &  \nodata  &  47/1     &  GG495   &  Loral 1K$\times$1K \\
2007 Apr 26       &  $5650-6790$  &  1700   & \phn 5  &  CTIO 1.5m + Cassegrain  &  \nodata  &  47/1     &  GG495   &  Loral 1K$\times$1K \\
2007 Jun 9        &  $5650-6790$  &  1700   & \phn 1  &  CTIO 1.5m + Cassegrain  &  \nodata  &  47/1     &  GG495   &  Loral 1K$\times$1K \\
2007 Jun 30       &  $5650-6790$  &  1700   & \phn 9  &  CTIO 1.5m + Cassegrain  &  \nodata  &  47/1     &  GG495   &  Loral 1K$\times$1K \\
2007 Jul 3        &  $5650-6790$  &  1700   & \phn 2  &  CTIO 1.5m + Cassegrain  &  \nodata  &  47/1     &  GG495   &  Loral 1K$\times$1K \\
2007 Jul $27-28$  &  $5650-6790$  &  1700   & \phn 7  &  CTIO 1.5m + Cassegrain  &  \nodata  &  47/1     &  GG495   &  Loral 1K$\times$1K \\
\enddata
\end{deluxetable}

% Table of Ha equivalent widths

\begin{deluxetable}{cccccccccc}
\tablewidth{0pt}
\tabletypesize{\scriptsize}
\tablecaption{H$\alpha$ Equivalent Widths\label{eqwidth}}
\tablehead{
\colhead{MG} &
\colhead{WEBDA} &
\colhead{$W_\lambda$ (\AA)} &
\colhead{$W_\lambda$ (\AA)} &
\colhead{$W_\lambda$ (\AA)} &
\colhead{$W_\lambda$ (\AA)} &
\colhead{$W_\lambda$ (\AA)} &
\colhead{$W_\lambda$ (\AA)} &
\colhead{$W_\lambda$ (\AA)} &
\colhead{$W_\lambda$ (\AA)} \\
\colhead{ID} &
\colhead{ID} &
\colhead{(2003 Mar)} &
\colhead{(2005 Feb)} &
\colhead{(2006 May)} &
\colhead{(2007 Jan/Feb)} &
\colhead{(2007 Apr)} &
\colhead{(2007 May)} &
\colhead{(2007 June)} &
\colhead{(2007 July)}
}
\startdata
%      MG ID  Webda       Mar 2003         Feb 2005         May 2006         Jan/Feb 2007	April 2007	May 2007	June 2007	July 2007

\phn\phn  2  &   \nodata  & \nodata        & \phn\phs 3.24  & \phn\phs 4.07  & \nodata          & \nodata   & \phn\phs 3.93  &  \nodata        &  \nodata        \\
    \phn 16  &       169  & \nodata        & \phn\phs 4.28  & \phn\phs 4.26  & \nodata          & \nodata   & \phn\phs 3.84  &  \nodata        &  \nodata        \\
    \phn 23  &   \nodata  & \nodata        & \phn\phs 9.22  & \phn\phs 9.37  & \nodata          & \nodata   & \phn\phs 7.59  &  \nodata        &  \nodata        \\
    \phn 25  &       291  & \phn\phs 3.62  & \phn\phs 3.18  & \phn\phs 5.30  & $-$1.63, $-$1.06 &    0.42   & \phn\phs 1.04  &  \phn\phs 1.42  &  \nodata        \\ % Be var
    \phn 27  &       146  & \phn\phs 5.14  & \phn\phs 5.38  & \phn\phs 5.69  & \nodata          & \nodata   & \phn\phs 4.74  &  \nodata        &  \nodata        \\
    \phn 31  &       151  & \phn\phs 2.91  & \phn\phs 3.99  & \phn\phs 0.57  & $-$3.34, $-$5.94 &  $-$6.64  & \phn  $-$6.80  &  \nodata        &  \nodata        \\ % Be var
    \phn 36  & \phn   13  & \nodata        & \phn\phs 8.46  & \phn\phs 8.56  & \nodata          & \nodata   & \phn\phs 6.93  &  \nodata        &  \nodata        \\
    \phn 41  &       130  & \nodata        & \phn\phs 4.88  & \phn\phs 5.28  & \nodata          & \nodata   & \phn\phs 4.41  &  \nodata        &  \nodata        \\
    \phn 42  &       178  & \nodata        & \phn\phs 6.07  & \phn\phs 6.02  & \nodata          & \nodata   & \phn\phs 5.22  &  \nodata        &  \nodata        \\
    \phn 45  & \phn\phn8  & \phn\phs 4.69  & \phn\phs 4.98  & \phn\phs 5.10  & \nodata          & \nodata   & \phn\phs 4.35  &  \nodata        &  \nodata        \\
    \phn 47  & \phn   15  &      $-$14.99  &      $-$11.33  &      $-$11.29  & \phn   $-$8.11   & \nodata   & \phs  $-$6.81  &  \phn  $-$5.61  &  \phn  $-$4.68  \\ % Be
    \phn 49  &       137  & \phn\phs 4.87  & \phn\phs 4.37  & \phn\phs 4.89  & \nodata          & \nodata   & \phn\phs 4.25  &  \nodata        &  \nodata        \\
    \phn 54  &       125  & \nodata        & \phn\phs 5.18  & \phn\phs 5.61  & \nodata          & \nodata   & \phn\phs 4.84  &  \nodata        &  \nodata        \\
    \phn 55  &   \nodata  & \nodata        & \phn\phs 5.03  & \phn\phs 5.12  & \nodata          & \nodata   & \phn\phs 4.59  &  \nodata        &  \nodata        \\
    \phn 57  & \phn   24  & \nodata        & \phn\phs 5.67  & \phn\phs 5.86  & \nodata          & \nodata   & \phn\phs 4.72  &  \nodata        &  \nodata        \\
    \phn 61  & \phn   20  & \nodata        & \phn\phs 4.09  & \phn\phs 3.54  & \phn\phs  3.66   & \nodata   & \phn\phs 3.56  &  \phn\phs 4.41  &  \phn\phs 1.20  \\
    \phn 69  &   \nodata  & \nodata        & \phn\phs 2.89  & \phn\phs 3.19  & \nodata          & \nodata   & \phn\phs 2.81  &  \nodata        &  \nodata        \\
    \phn 72  & \phn\phn4  & \phn\phs 7.19  & \phn\phs 8.27  & \phn\phs 8.65  & \phn\phs  7.14   & \nodata   & \phn\phs 7.08  &  \nodata        &  \phn\phs 7.25  \\
    \phn 73  & \phn   26  & \phn\phs 3.65  & \phn\phs 0.42  & \phn\phs 3.28  & \phn\phs  1.27   & \nodata   & \phn  $-$0.35  &  \nodata        &  \phn  $-$0.92  \\ % Be var
    \phn 77  &       195  & \phn\phs 4.85  & \phn\phs 4.61  & \phn\phs 5.72  & \nodata          & \nodata   & \phn\phs 4.73  &  \nodata        &  \nodata        \\
    \phn 83  & \phn   27  & \phn\phs 0.53  & \phn\phs 4.07  & \phn\phs 0.84  & \phn\phs  2.92   &    3.72   & \phn\phs 2.93  &  \nodata        &  \nodata        \\ % Be var
    \phn 92  & \phn\phn1  & \phn\phs 2.06  & \phn\phs 2.31  & \phn\phs 2.73  & \phn\phs  3.32   & \nodata   & \phn\phs 3.41  &  \phn\phs 2.66  &  \nodata        \\ % Be
    \phn 94  &       194  & \nodata        & \phn\phs 5.94  & \phn\phs 5.91  & \nodata          & \nodata   & \phn\phs 4.94  &  \nodata        &  \nodata        \\
    \phn 96  & \phn   45  & \nodata        & \phn\phs 6.39  & \phn\phs 6.87  & \nodata          & \nodata   & \phn\phs 6.26  &  \nodata        &  \nodata        \\
    \phn 98  & \phn   36  & \phn\phs 4.99  & \phn\phs 3.10  & \phn\phs 0.21  & \phn\phs  4.43   &    4.36   & \phn\phs 4.14  &  \nodata        &  \phn\phs 5.05  \\ % Be var
        101  & \phn   34  & \nodata        & \phn\phs 7.40  & \phn\phs 7.44  & \nodata          & \nodata   & \nodata        &  \nodata        &  \nodata        \\
        118  & \nodata    & \nodata        & \phn\phs 7.40  & \phn\phs 7.03  & \nodata          & \nodata   & \phn\phs 5.73  &  \nodata        &  \nodata        \\
        119  & \phn   81  & \phn\phs 4.76  & \phn  $-$5.48  & \phn  $-$3.19  & \phn  $-$6.53    & \nodata   & \phn  $-$8.67  &  \nodata        &       $-$12.83  \\ % Be var
        126  & \nodata    & \nodata        & \phn\phs 7.46  & \phn\phs 8.49  & \nodata          & \nodata   & \phn\phs 6.86  &  \nodata        &  \nodata        \\
        127  & \phn   53  & \phn  $-$7.21  & \phn  $-$4.95  & \phn  $-$4.00  & \phn  $-$4.19    & \nodata   & \phn  $-$5.47  &  \phn  $-$5.60  &       $-$12.81  \\ % Be
        129  & \nodata    & \nodata        & \phn\phs 8.66  & \phn\phs 7.94  & \nodata          & \nodata   & \phn\phs 6.84  &  \nodata        &  \nodata        \\
        130  & \phn   67  & \nodata        & \phn\phs 2.00  & \phn\phs 5.04  & \phn\phs 4.23    & \nodata   & \phn\phs 4.44  &  \phn\phs 4.38  &  \nodata        \\ % new Be
        133  & \phn   63  & \phn\phs 4.40  & \phn  $-$1.15  & \phn  $-$9.14  &      $-$11.62    & \nodata   &      $-$12.72  &       $-$12.40  &  \nodata        \\ % Be var
        139  &       204  & \phn\phs 8.48  & \phn\phs 5.46  & \phn\phs 4.24  & \phn\phs 4.51    &    4.50   & \phn\phs 4.42  &  \phn\phs 5.54  &  \nodata        \\ % Be var
        154  & \phn   88  &      $-$33.46  &      $-$40.07  &      $-$33.56  &      $-$34.37    & \nodata   &      $-$33.95  &  \nodata        &       $-$36.16  \\ % Be
        155  & \nodata    & \nodata        & \phn\phs 7.92  & \phn\phs 7.90  & \nodata          & \nodata   & \phn\phs 6.38  &  \nodata        &  \nodata        \\
        161  & \phn   70  & \nodata        & \phn\phs 3.92  & \phn\phs 4.19  & \nodata          & \nodata   & \phn\phs 3.61  &  \nodata        &  \nodata        \\
        162  & \nodata    & \nodata        & \phn\phs 9.15  & \phn\phs 9.23  & \nodata          & \nodata   & \phn\phs 7.45  &  \nodata        &  \nodata        \\
        170  & \phn   94  & \nodata        & \phn\phs 4.81  & \phn\phs 5.26  & \nodata          & \nodata   & \phn\phs 4.17  &  \nodata        &  \nodata        \\
        173  &       233  & \nodata        & \phn\phs 7.79  & \phn\phs 7.92  & \nodata          & \nodata   & \phn\phs 6.62  &  \nodata        &  \nodata        \\
        175  &       218  & \nodata        & \phn\phs 6.67  & \phn\phs 6.74  & \nodata          & \nodata   & \phn\phs 6.01  &  \nodata        &  \nodata        \\
        178  &       213  & \nodata        & \phn\phs 6.32  & \phn\phs 7.71  & \nodata          & \nodata   & \phn\phs 6.35  &  \nodata        &  \nodata        \\
        190  &       260  & \nodata        & \phn\phs 6.81  & \phn\phs 6.86  & \nodata          & \nodata   & \nodata        &  \nodata        &  \nodata        \\
        196  &       239  & \phn\phs 4.52  & \phn\phs 4.32  & \phn\phs 4.39  & \nodata          & \nodata   & \phn\phs 1.46  &  \nodata        &  \nodata        \\
        197  &       253  & \nodata        & \phn\phs 7.65  & \phn\phs 8.85  & \phn\phs 7.04    & \nodata   & \phn\phs 7.09  &  \nodata        &  \phn\phs 7.66  \\
        198  &       264  &      $-$39.49  &      $-$43.40  &      $-$53.71  &      $-$52.77    & \nodata   &      $-$53.44  &       $-$54.69  &  \nodata        \\ % Be
        200  &       240  & \phn  $-$8.84  & \phn  $-$6.66  & \phn  $-$4.62  & \phn  $-$4.88    & \nodata   & \phn  $-$5.19  &  \phn  $-$5.76  &  \nodata        \\ % Be
\enddata
\end{deluxetable}

\begin{deluxetable}{llcccccccccccc}
\tablewidth{0pt}
\tabletypesize{\scriptsize}
\tablecaption{Physical Parameters of Cluster Members\label{params} }
\tablehead{
\colhead{ } &
\colhead{MG } &
\colhead{$V \sin i$} &
\colhead{$\Delta V \sin i$} &
\colhead{$T_{\rm eff}$} &
\colhead{$\Delta T_{\rm eff}$} &
\colhead{ } &
\colhead{ } &
\colhead{ } &
\colhead{$M_\star$} &
\colhead{$\Delta M_\star$} &
\colhead{$R_\star$} &
\colhead{$\Delta R_\star$} &
\colhead{$V_{\rm crit}$} \\
\colhead{ } &
\colhead{ID} &
\colhead{(km~s$^{-1}$)} &
\colhead{(km~s$^{-1}$)} &
\colhead{(K)} &
\colhead{(K)} &
\colhead{$\log g$} &
\colhead{$\Delta \log g$} &
\colhead{$\log g_{\rm polar}$} &
\colhead{($M_\odot$)} &
\colhead{($M_\odot$)} &
\colhead{($R_\odot$)} &
\colhead{($R_\odot$)} &
\colhead{(km~s$^{-1}$)} }
\startdata
B stars: 
 & \phn\phn  2\tablenotemark{a}  &     307  & \phn 5  &  17071  &  \phn450  &  3.10  &  0.05  &  3.57  &   7.0  &   0.3  &   \phn 7.2  &   0.6  &  352 \\
 & \phn     16\tablenotemark{a}  &     144  & \phn 7  &  19420  &  \phn300  &  3.48  &  0.02  &  3.62  &   8.6  &   0.2  &   \phn 7.5  &   0.5  &  380 \\
 & \phn     27\tablenotemark{a}  &     287  & \phn 6  &  16033  &  \phn200  &  3.83  &  0.02  &  4.10  &   4.9  &   0.1  &   \phn 3.3  &   0.8  &  436 \\
 & \phn     36\tablenotemark{a}  &     283  &     22  &  11931  &  \phn145  &  4.02  &  0.05  &  4.27  &   3.0  &   0.0  &   \phn 2.1  &   0.1  &  424 \\
 & \phn     41\tablenotemark{a}  & \phn 84  & \phn 7  &  17900  &  \phn300  &  3.78  &  0.03  &  3.84  &   6.6  &   0.1  &   \phn 5.1  &   0.8  &  406 \\
 & \phn     42\tablenotemark{a}  &     239  & \phn 5  &  16100  &  \phn100  &  3.90  &  0.02  &  4.11  &   4.9  &   0.0  &   \phn 3.2  &   0.8  &  438 \\
% & \phn     45\tablenotemark{c}  &  \nodata  &  \nodata  &  19332  &  \phn301  &  3.92  &  0.18  &  \nodata  &   \nodata  &   \nodata  &   \nodata  &   \nodata  &  \nodata \\
 & \phn     49\tablenotemark{a}  &     252  & \phn 9  &  17400  &  \phn100  &  3.75  &  0.02  &  3.99  &   6.0  &   0.1  &   \phn 4.1  &   0.9  &  431 \\
 & \phn     54\tablenotemark{a}  &     192  & \phn 9  &  17340  &  \phn250  &  3.82  &  0.02  &  3.99  &   5.9  &   0.1  &   \phn 4.1  &   0.9  &  430 \\
 & \phn     55\tablenotemark{a}  &     121  & \phn 6  &  18000  &  \phn200  &  3.78  &  0.03  &  3.88  &   6.6  &   0.1  &   \phn 4.9  &   0.8  &  413 \\
 & \phn     57\tablenotemark{a}  &     238  & \phn 8  &  16760  &  \phn250  &  3.98  &  0.03  &  4.17  &   5.1  &   0.2  &   \phn 3.1  &   1.0  &  459 \\
 & \phn     61\tablenotemark{a, b}  &     331  & \phn 8  &  18883  &  \phn350  &  3.23  &  0.02  &  3.71  &   7.7  &   0.3  &   \phn 6.4  &   0.6  &  390 \\  % SB2?
% & \phn     72\tablenotemark{c}  &  \nodata  &  \nodata  &  19152  &  \phn301  &  3.50  &  0.18  &  \nodata  &   \nodata  &   \nodata  &   \nodata  &   \nodata  &  \nodata \\
 & \phn     77\tablenotemark{a}  &     338  &     10  &  16000  &  \phn200  &  3.80  &  0.05  &  4.15  &   4.8  &   0.1  &   \phn 3.1  &   0.9  &  445 \\
 & \phn     94\tablenotemark{a}  &     177  & \phn 5  &  15650  &  \phn150  &  3.78  &  0.03  &  3.94  &   5.0  &   0.1  &   \phn 4.0  &   0.5  &  401 \\
 & \phn     96\tablenotemark{a}  &     121  & \phn 5  &  13937  &  \phn171  &  4.31  &  0.04  &  4.37  &   3.6  &   0.1  &   \phn 2.0  &   0.3  &  472 \\
 &         101\tablenotemark{a}  &     309  &     14  &  12753  &  \phn192  &  3.84  &  0.05  &  4.14  &   3.5  &   0.1  &   \phn 2.6  &   0.4  &  409 \\
 &         118\tablenotemark{a}  &     242  & \phn 9  &  13353  &  \phn191  &  3.89  &  0.05  &  4.11  &   3.7  &   0.1  &   \phn 2.8  &   0.5  &  409 \\
 &         126\tablenotemark{a}  & \phn 83  & \phn 9  &  12399  &  \phn318  &  3.79  &  0.09  &  3.87  &   3.8  &   0.1  &   \phn 3.7  &   0.4  &  358 \\
 &         129\tablenotemark{a}  & \phn 97  &     27  &  12034  &  \phn203  &  3.77  &  0.06  &  3.87  &   3.7  &   0.1  &   \phn 3.7  &   0.3  &  355 \\
 &         155\tablenotemark{a}  &     231  &     11  &  12694  &  \phn153  &  4.05  &  0.04  &  4.22  &   3.3  &   0.1  &   \phn 2.3  &   0.5  &  425 \\
 &         161\tablenotemark{a}  & \phn 73  & \phn 7  &  18400  &  \phn250  &  3.43  &  0.02  &  3.49  &   8.3  &   0.2  &   \phn 8.5  &   0.4  &  351 \\
 &         162\tablenotemark{a}  &     194  &     11  &  11513  &  \phn148  &  3.99  &  0.05  &  4.12  &   3.0  &   0.1  &   \phn 2.5  &   0.2  &  390 \\
 &         170\tablenotemark{a}  & \phn 65  & \phn 6  &  18060  &  \phn250  &  3.78  &  0.03  &  3.82  &   6.8  &   0.1  &   \phn 5.3  &   0.7  &  403 \\
 &         173\tablenotemark{a}  &     171  & \phn 7  &  14210  &  \phn192  &  4.21  &  0.04  &  4.30  &   3.8  &   0.1  &   \phn 2.3  &   0.1  &  459 \\
 &         175\tablenotemark{a}  &     252  &     13  &  13823  &  \phn193  &  4.01  &  0.05  &  4.21  &   3.8  &   0.1  &   \phn 2.5  &   0.6  &  435 \\
 &         178\tablenotemark{a}  &     160  &     12  &  12504  &  \phn141  &  3.92  &  0.04  &  4.05  &   3.5  &   0.1  &   \phn 2.9  &   0.3  &  390 \\
 &         197\tablenotemark{a, b}  &     198  &     15  &  12402  &  \phn182  &  4.15  &  0.06  &  4.25  &   3.2  &   0.1  &   \phn 2.2  &   0.5  &  427 \\  % SB2?
\\
Be stars: 
 & \phn     25\tablenotemark{c}  &     261  & \phn 6  &  18995  &  \phn301  &  4.02  &  0.09  &  4.02  &   6.6  &   0.1  &   \phn 4.1  &   0.9  &  450 \\ % Be
 & \phn     31\tablenotemark{d}  &     197  & \phn 5  &  17834  &  \phn400  &  3.82  &  0.10  &  3.99  &   6.2  &   0.2  &   \phn 4.2  &   1.0  &  435 \\ % Be
 & \phn     47\tablenotemark{c}  &     190  & \phn 5  &  18399  &  \phn301  &  3.30  &  0.09  &  3.30  &   9.3  &   0.1  &   11.3  &   2.3  &  323 \\ % Be
 & \phn     73\tablenotemark{c}  &     296  & \phn 5  &  18274  &  \phn301  &  3.49  &  0.09  &  3.49  &   8.2  &   0.1  &   \phn 8.5  &   1.8  &  349 \\ % Be
 & \phn     83\tablenotemark{c}  &   \nodata  &  \nodata  &  18817  &  \phn301  &  3.31  &  0.09  &  3.31  &   9.9  &   0.1  &   11.5  &   2.4  & 329  \\
 & \phn     92\tablenotemark{c}  &     214  &     17  &  18725  &  \phn301  &  3.34  &  0.09  &  3.34  &   9.5  &   0.1  &   10.8  &   2.3  &  332 \\ % Be
 & \phn     98\tablenotemark{c}  &   \nodata  &  \nodata  &  16890  &  \phn301  &  3.84  &  0.10  &  3.84 &  6.2  &   0.1  &   \phn 4.9  &   1.0  & 398 \\
 &         119\tablenotemark{c}  &   \nodata  &  \nodata  &  17792  &  \phn301  &  3.75  &  0.09  &  3.75  &   6.8  &   0.1  &  \phn 5.8  &   1.2  &  387 \\
 &         127\tablenotemark{d}  &     347  & \phn 5  &  17687  &  \phn550  &  3.61  &  0.12  &  4.01  &   6.1  &   0.3  &   \phn 4.0  &   1.0  &  437 \\ % Be
 &         130\tablenotemark{c}  &     285  & \phn 7  &  17519  &  \phn301  &  3.69  &  0.09  &  3.69  &   6.9  &   0.1  &   \phn 6.2  &   1.3  &  375 \\  % Be
 &         133\tablenotemark{c}  &   \nodata  &  \nodata  &  18564  &  \phn301  &  3.53  &  0.09  &  3.53  &   8.3  &   0.1  &   \phn 8.1  &   1.7  &  358 \\
 &         139\tablenotemark{c}  &     343  &     12  &  15945  &  \phn301  &  3.95  &  0.10  &  3.95  &   5.2  &   0.1  &   \phn 4.0  &   0.8  &  406 \\  % Be
 &         154\tablenotemark{c}  &     112  &     29  &  13254  &  1254  &  3.29  &  0.18  &  3.29  &   5.6  &  0.3   &   \phn 8.8  &   1.8  &  282 \\  % Be, shell
 &         196\tablenotemark{a}  &     165  & \phn 5  &  19660  &  \phn250  &  3.73  &  0.02  &  3.87  &   7.6  &   0.2  &   \phn 5.3  &   1.1  &  426 \\
 &         198\tablenotemark{c}  &     251  & \phn 6  &  19580  &  \phn301  &  4.00  &  0.09  &  4.00  &   7.0  &   0.1  &   \phn 4.4  &   0.9  &   449 \\ % Be
 &         200\tablenotemark{c}  &     236  &     12  &  16301  &  \phn301  &  3.51  &  0.10  &  3.51  &   6.8  &   0.1  &   \phn 7.5  &   1.6  &  337  \\  % Be
\enddata
\tablenotetext{a}{$T_{\rm eff}$ and $\log g$ measured from H$\gamma$ line fit.}
\tablenotetext{b}{Parameters are unreliable since the star is a suspected SB2.}
\tablenotetext{c}{$T_{\rm eff}$ and $\log g$ measured from Str\"omgren photometry.}
\tablenotetext{d}{$T_{\rm eff}$ and $\log g$ measured from \ion{He}{1} line fits.}
\end{deluxetable}

\begin{deluxetable}{llcccccc}
\tablewidth{0pt}
\tabletypesize{\scriptsize}
\tablecaption{Estimated Sizes of Be Star Disks\label{bevar} }
\tablehead{
\colhead{MG} &
\colhead{HJD$-$} &
\colhead{Assumed} &
\colhead{$i$} &
\colhead{$\log \rho_0$} &
\colhead{ } &
\colhead{$M_{\rm disk}$} &
\colhead{$\Delta M_{\rm disk}/\Delta t$} \\
\colhead{ID} &
\colhead{2,450,000} &
\colhead{$V/V_{\rm crit}$} &
\colhead{(deg)} &
\colhead{(g~cm$^{-3}$)} &
\colhead{$R_{\rm disk}/R_{\star}$} &
\colhead{($10^{-11} M_\odot$)} &
\colhead{($10^{-11} M_\odot$~yr$^{-1}$)} }
\startdata
\phn 25 &  2720.825              &   0.7   &   64.9  &  $-$12.2  &  2.3  & \phn    41.9  & \nodata           \\
        &  3403.79          \phn & \nodata & \nodata &  $-$12.1  &  2.5  & \phn    44.3  & \phn\phn\phs 1.3  \\
        &  3870 \phn\phn\phn\phn & \nodata & \nodata &  $-$12.3  &  1.6  & \phn    30.6  & \phn     $-$10.7  \\
        &  4120.664              & \nodata & \nodata &  $-$12.0  &  3.8  & \phn    66.1  & \phn\phs    51.7  \\
        &  4129.743              & \nodata & \nodata &  $-$12.0  &  3.7  & \phn    63.7  & \phn     $-$92.7  \\
        &  4216.493              & \nodata & \nodata &  $-$12.0  &  3.3  & \phn    57.5  & \phn     $-$26.3  \\
        &  4225.66          \phn & \nodata & \nodata &  $-$12.1  &  3.1  & \phn    54.7  &         $-$110.0  \\
        &  4260.635              & \nodata & \nodata &  $-$12.1  &  3.0  & \phn    53.0  & \phn     $-$17.8  \\
\\
\phn 31 &  2720.744              &   0.7   &   40.5  &  $-$12.1  &  2.5  & \phn    19.8  & \nodata           \\
        &  3403.79          \phn & \nodata & \nodata &  $-$12.2  &  2.1  & \phn    17.2  & \phn\phn  $-$1.4  \\
        &  3870 \phn\phn\phn\phn & \nodata & \nodata &  $-$12.0  &  3.0  & \phn    24.6  & \phn\phn\phs 5.8  \\
        &  4120.677              & \nodata & \nodata &  $-$11.9  &  3.7  & \phn    31.4  & \phn\phn\phs 9.9  \\
        &  4129.771              & \nodata & \nodata &  $-$11.9  &  4.1  & \phn    35.0  & \phs       142.4  \\
        &  4216.512              & \nodata & \nodata &  $-$11.9  &  4.1  & \phn    35.7  & \phn\phn\phs 3.3  \\
        &  4225.66          \phn & \nodata & \nodata &  $-$11.9  &  4.2  & \phn    35.9  & \phn\phn\phs 6.7  \\
\\
\phn 47 &  2719.557              &   0.7   &   40.8  &  $-$11.8  &  4.5  & \phn    72.3  & \nodata           \\
        &  3403.79          \phn & \nodata & \nodata &  $-$11.8  &  4.4  & \phn    70.3  & \phn\phn  $-$1.1  \\
        &  3870 \phn\phn\phn\phn & \nodata & \nodata &  $-$11.8  &  4.4  & \phn    70.2  & \phn\phn\phs 0.0  \\
        &  4133.656              & \nodata & \nodata &  $-$11.9  &  4.2  & \phn    66.3  & \phn\phn  $-$5.4  \\
        &  4225.66          \phs & \nodata & \nodata &  $-$11.9  &  4.1  & \phn    64.1  & \phn\phn  $-$8.8  \\
        &  4281.495              & \nodata & \nodata &  $-$11.9  &  4.0  & \phn    61.7  & \phn     $-$15.7  \\
        &  4309.466              & \nodata & \nodata &  $-$11.9  &  3.9  & \phn    59.5  & \phn     $-$28.5  \\
\\
\phn 73 &  2719.567              &   0.7   &   73.1  &  $-$12.2  &  2.5  & \phn    16.7  & \nodata           \\
        &  3403.79          \phn & \nodata & \nodata &  $-$12.0  &  3.5  & \phn    23.0  & \phn\phn\phs 3.4  \\
        &  3870 \phn\phn\phn\phn & \nodata & \nodata &  $-$12.1  &  2.6  & \phn    17.5  & \phn\phn  $-$4.3  \\
        &  4120.711              & \nodata & \nodata &  $-$12.1  &  3.3  & \phn    21.5  & \phn\phn\phs 5.8  \\
        &  4225.66          \phn & \nodata & \nodata &  $-$12.0  &  3.7  & \phn    24.3  & \phn\phn\phs 9.9  \\
        &  4308.554              & \nodata & \nodata &  $-$12.0  &  3.9  & \phn    25.3  & \phn\phn\phs 4.1  \\
\\
\phn 92 &  2719.629              &   0.7   &   48.9  &  $-$12.1  &  2.7  & \phn    50.4  & \nodata           \\
        &  3403.79          \phn & \nodata & \nodata &  $-$12.1  &  2.6  & \phn    49.1  & \phn\phn  $-$0.7  \\
        &  3870 \phn\phn\phn\phn & \nodata & \nodata &  $-$12.1  &  2.5  & \phn    47.0  & \phn\phn  $-$1.7  \\
        &  4133.668              & \nodata & \nodata &  $-$12.1  &  2.3  & \phn    43.9  & \phn\phn  $-$4.3  \\
        &  4225.66          \phn & \nodata & \nodata &  $-$12.2  &  2.3  & \phn    43.4  & \phn\phn  $-$1.9  \\
        &  4281.510              & \nodata & \nodata &  $-$12.1  &  2.5  & \phn    47.3  & \phn\phs    25.9  \\
\\
    127 &  2719.759              &   0.8   &   80.1  &  $-$11.9  &  5.6  & \phn    30.7  & \nodata           \\
        &  3403.79          \phn & \nodata & \nodata &  $-$11.9  &  5.2  & \phn    28.5  & \phn\phn  $-$1.2  \\
        &  3870 \phn\phn\phn\phn & \nodata & \nodata &  $-$11.9  &  5.0  & \phn    27.4  & \phn\phn  $-$0.9  \\
        &  4133.680              & \nodata & \nodata &  $-$11.9  &  5.0  & \phn    27.6  & \phn\phn\phs 0.3  \\
        &  4225.66          \phn & \nodata & \nodata &  $-$11.9  &  5.3  & \phn    29.1  & \phn\phn\phs 5.8  \\
        &  4281.524              & \nodata & \nodata &  $-$11.9  &  5.3  & \phn    29.2  & \phn\phn\phs 0.9  \\
        &  4309.495              & \nodata & \nodata &  $-$11.8  &  6.1  & \phn    33.8  & \phn\phs    59.9  \\
\\
    130 &  3403.79          \phn &   0.7   &   63.5  &  $-$12.1  &  3.0  & \phn    11.9  & \nodata           \\
        &  3870 \phn\phn\phn\phn & \nodata & \nodata &  $-$12.3  &  1.8  & \phn\phn 7.7  & \phn\phn  $-$3.3  \\
        &  4120.750              & \nodata & \nodata &  $-$12.2  &  2.2  & \phn\phn 9.0  & \phn\phn\phs 1.9  \\
        &  4225.66          \phn & \nodata & \nodata &  $-$12.2  &  2.1  & \phn\phn 8.7  & \phn\phn  $-$1.1  \\
        &  4281.540              & \nodata & \nodata &  $-$12.2  &  2.1  & \phn\phn 8.8  & \phn\phn\phs 0.6  \\
\\
    139 &  2720.796              &   0.9   &   50.6  &  \nodata  &  0.0  & \phn\phn 0.0  & \nodata           \\
        &  3403.79          \phn & \nodata & \nodata &  $-$12.3  &  1.8  & \phn\phn 1.7  & \phn\phn\phs 0.9  \\
        &  3870 \phn\phn\phn\phn & \nodata & \nodata &  $-$12.2  &  2.3  & \phn\phn 2.2  & \phn\phn\phs 0.4  \\
        &  4120.770              & \nodata & \nodata &  $-$12.2  &  2.2  & \phn\phn 2.1  & \phn\phn  $-$0.1  \\
        &  4216.632              & \nodata & \nodata &  $-$12.2  &  2.2  & \phn\phn 2.2  & \phn\phn\phs 0.0  \\
        &  4225.66          \phn & \nodata & \nodata &  $-$12.2  &  2.3  & \phn\phn 2.2  & \phn\phn\phs 0.0  \\
        &  4281.572              & \nodata & \nodata &  $-$12.3  &  1.8  & \phn\phn 1.7  & \phn\phn  $-$3.2  \\
\\
    196 &  3870 \phn\phn\phn\phn &   0.7   &   33.5  &  \nodata  &  0.0  & \phn\phn 0.0  & \nodata           \\
        &  4225.66          \phn & \nodata & \nodata &  $-$12.1  &  2.6  & \phn    15.0  & \phn\phs    15.4  \\
\\
    198 &  2720.658              &   0.7   &   69.9  &  $-$11.9  &  5.0  &        253.4  & \nodata           \\
        &  3403.79          \phn & \nodata & \nodata &  $-$11.9  &  4.9  &        248.9  & \phn\phn  $-$2.4  \\
        &  3870 \phn\phn\phn\phn & \nodata & \nodata &  $-$11.9  &  4.7  &        238.7  & \phn\phn  $-$8.0  \\
        &  4133.732              & \nodata & \nodata &  $-$11.9  &  4.7  &        239.5  & \phn\phn\phs 1.1  \\
        &  4225.66          \phn & \nodata & \nodata &  $-$11.9  &  4.7  &        238.9  & \phn\phn  $-$2.4  \\
        &  4281.589              & \nodata & \nodata &  $-$11.9  &  4.7  &        237.8  & \phn\phn  $-$7.1  \\
\\
    200 &  2720.637              &   0.7   &   46.2  &  $-$11.8  &  4.8  & \phn\phn 9.9  & \nodata           \\
        &  3403.79          \phn & \nodata & \nodata &  $-$11.9  &  4.6  & \phn\phn 9.4  & \phn\phn  $-$0.3  \\
        &  3870 \phn\phn\phn\phn & \nodata & \nodata &  $-$11.9  &  4.3  & \phn\phn 8.7  & \phn\phn  $-$0.5  \\
        &  4133.741              & \nodata & \nodata &  $-$11.9  &  4.3  & \phn\phn 8.8  & \phn\phn\phs 0.1  \\
        &  4225.66          \phn & \nodata & \nodata &  $-$11.9  &  4.4  & \phn\phn 8.9  & \phn\phn\phs 0.4  \\
        &  4281.599              & \nodata & \nodata &  $-$11.9  &  4.5  & \phn\phn 9.1  & \phn\phn\phs 1.2  \\
\enddata
\end{deluxetable}

\end{document}